%% file: main.tex
\documentclass[camera,10pt]{jpaper} 
\usepackage{titlesec}
\titlespacing\section{0pt}{*0.5}{*0.25}
\titlespacing\subsection{0pt}{*0.5}{*0.25}
\titlespacing\subsubsection{0pt}{*0.5}{*0.25}

\usepackage[utf8]{inputenc}

\DeclareMathAlphabet{\mathcal}{OMS}{cmsy}{m}{n}
\usepackage{setspace} 
\usepackage[italic]{mathastext}
\usepackage{array}

\newcommand{\ignore}[1]{}
\usepackage{dblfloatfix}
\usepackage{fancyhdr}
\usepackage[normalem]{ulem}
\usepackage{datetime} 
\usepackage[hyphens]{url}
\usepackage{multirow}
\usepackage{multicol}
\usepackage{float}
\usepackage[shortlabels]{enumitem}
\usepackage{bm}
\usepackage[sort,compress]{cite}
\usepackage{tikz}
\usepackage{xcolor}
\usepackage[font={small,bf}]{subcaption}
\usepackage[linesnumbered,ruled]{algorithm2e}
\usepackage{balance} 

\def\BibTeX{{\rm B\kern-.05em{\sc i\kern-.025em b}\kern-.08em
    T\kern-.1667em\lower.7ex\hbox{E}\kern-.125emX}}

\usepackage[font={small,bf}]{caption}
\setlist[itemize]{leftmargin=*}

\usepackage{pifont}
\usepackage{amssymb}
\usepackage{amsmath}

\usepackage[bookmarks=true,breaklinks=true,letterpaper=true,colorlinks,linkcolor=black,citecolor=black,urlcolor=black]{hyperref}

\definecolor{dollarbill}{rgb}{0.52, 0.73, 0.4}

\newcolumntype{L}[1]{>{\raggedright\let\newline\\\arraybackslash\hspace{0pt}}m{#1}}
\newcolumntype{C}[1]{>{\centering\let\newline\\\arraybackslash\hspace{0pt}}m{#1}}
\newcolumntype{R}[1]{>{\raggedleft\let\newline\\\arraybackslash\hspace{0pt}}m{#1}}

\setlist[itemize]{noitemsep,itemsep=0pt,parsep=0pt,topsep=0pt,partopsep=0pt,leftmargin=1em}
\setlist[enumerate]{noitemsep,itemsep=0pt,parsep=0pt,topsep=0pt,partopsep=0pt,leftmargin=1em}



\definecolor{amber}{rgb}{1.0, 0.49, 0.0}
\definecolor{darkamber}{rgb}{0.9, 0.49, 0.0}
\definecolor{darkgreen}{rgb}{0.0, 0.2, 0.13}
\definecolor{darkbyzantium}{rgb}{0.36, 0.22, 0.33}
\definecolor{darkseagreen}{rgb}{0.56, 0.74, 0.56}
\definecolor{darkspringgreen}{rgb}{0.09, 0.45, 0.27}
\definecolor{dollarbill}{rgb}{0.52, 0.73, 0.4}
\definecolor{darkcerulean}{rgb}{0.03, 0.27, 0.49}

\newif\ifcameraready
\camerareadytrue

\ifcameraready
  \usepackage[disable]{todonotes}

  \newcommand{\mpo}[1]{#1}
  \newcommand{\mpt}[1]{#1}
  \newcommand{\mpf}[1]{#1}
  
  \newcommand{\jk}[1]{#1}
  \newcommand{\mps}[1]{#1}
  \newcommand{\revmp}[1]{#1}
  
  \newcommand{\revmpf}[1]{#1}
  \newcommand{\cro}[1]{#1}
  \newcommand{\mpi}[1]{#1}
  \newcommand{\mpii}[1]{#1}
  \newcommand{\mpiii}[1]{#1}
  \newcommand{\mpiv}[1]{#1}
  \newcommand{\mpv}[1]{#1}
  \newcommand{\mpvi}[1]{#1}
  \newcommand{\mpvii}[1]{#1}
  \newcommand{\mpviii}[1]{#1}
  \newcommand{\hht}[1]{#1}
  \newcommand{\jkz}[1]{#1}
  \newcommand{\jkx}[1]{#1}
\else
  \usepackage[textsize=scriptsize]{todonotes}
  \makeatletter
    \renewcommand{\todo}[2][]{\@bsphack\@todo[#1]{\textcolor{black}{#2}}\@esphack\ignorespaces}
  \makeatother

  \newcommand{\mpo}[1]{#1}
  \newcommand{\mpt}[1]{#1}
  \newcommand{\mpf}[1]{#1}
  
  \newcommand{\jk}[1]{#1} 
  \newcommand{\mps}[1]{#1}
  \newcommand{\revmp}[1]{#1}
  
  \newcommand{\revmpf}[1]{#1}
  \newcommand{\cro}[1]{#1}
  
  \newcommand{\mpi}[1]{#1}
  \newcommand{\mpii}[1]{#1}
  \newcommand{\mpiii}[1]{#1}
  \newcommand{\mpiv}[1]{#1}
  \newcommand{\mpv}[1]{#1}
  \newcommand{\mpvi}[1]{#1}
  \newcommand{\mpvii}[1]{#1}
  \newcommand{\mpviii}[1]{#1}
  \newcommand{\hht}[1]{#1}
  \newcommand{\jkz}[1]{#1}
  \newcommand{\jkx}[1]{#1}
  \fi

\makeatletter
\g@addto@macro{\normalsize}{%
  \setlength{\abovedisplayskip}{2pt plus 1pt minus 1pt}
  \setlength{\belowdisplayskip}{2pt plus 1pt minus 1pt}
  \setlength{\abovedisplayshortskip}{0pt}
  \setlength{\belowdisplayshortskip}{0pt}
  \setlength{\intextsep}{2pt plus 1pt minus 1pt}
  \setlength{\textfloatsep}{4pt plus 1pt minus 1pt}
  \setlength{\skip\footins}{5pt plus 1pt minus 1pt}}
\makeatother

\makeatletter
\newcommand*{\textoverline}[1]{$\overline{\raisebox{0pt}[0.85\height]{#1}}\m@th$}
\makeatother

\makeatletter
\newcommand\requiredelimiter[2][########]{%
  \ifdefined#2%
    \def\@temp{\def#2#1}%
    \expandafter\@temp\expandafter{#2}%
  \else
    \@latex@error{\noexpand#2undefined}\@ehc
  \fi
}
\@onlypreamble\requiredelimiter
\makeatother

\newcommand*\circled[1]{\tikz[baseline=(char.base)]{
    \node[shape=circle,fill,inner sep=0.5pt,text width=5pt] (char) {\textcolor{white}{\textbf{#1}}};}}

\title{Bit-Exact ECC Recovery (BEER): \\ Determining DRAM On-Die ECC Functions
\\  by Exploiting DRAM Data Retention Characteristics}

\newcommand{\ethz}{{\large$^\dagger$}} 
\newcommand{\cmu}{{\large$^\ddagger$}}

\author{ \vspace{-2ex} \\
Minesh Patel\ethz \hspace{0.15in} 
Jeremie S. Kim\cmu\ethz \hspace{0.15in}
Taha Shahroodi\ethz \hspace{0.15in} 
Hasan Hassan\ethz \hspace{0.15in}
Onur Mutlu\ethz\cmu \vspace{2mm} \\
\textit{\ethz ETH Z{\"u}rich\hspace{0.2in} \cmu Carnegie Mellon University}
\\}

\begin{document}
\maketitle

\newcommand\blfootnote[1]{%
  \begingroup
  \renewcommand\thefootnote{}\footnotetext{#1}%
  \addtocounter{footnote}{-1}%
  \endgroup
}

\fancyhf{}
\fancypagestyle{firstpage}
{
  \renewcommand{\headrulewidth}{0pt}
  \pagenumbering{arabic}
  \fancyfoot[C]{\thepage}
}

\fancypagestyle{otherpagestyle}
{
  \renewcommand{\headrulewidth}{0pt}
  \pagenumbering{arabic}
  \fancyfoot[C]{\thepage}
}

\thispagestyle{firstpage}
\pagestyle{otherpagestyle}

\setstretch{0.9022}

\sloppypar
\begin{abstract}
  \input{0_abstract}
\end{abstract}

\input{1_introduction} 
\input{2_motivation} 
\setstretch{0.89}
\input{3_background}
\input{4_beer}
\input{5_beer}
\input{6_beer_eval}

\input{7_use_cases}
\input{8_related_work}
\input{9_conclusion}

\section*{Acknowledgments}
\mpii{We thank the SAFARI \mpiii{Research Group} members for the valuable input
and stimulating intellectual environment they provide, Karthik Sethuraman for
his expertise in nonparametric statistics, \mpviii{and the anonymous reviewers
for their feedback.}}


\clearpage
\balance
\bibliographystyle{IEEEtranS}
\bibliography{references}

\end{document}

%% file: 0_abstract.tex
Increasing single-cell DRAM error rates have pushed DRAM manufacturers to adopt
on-die error-\mpii{correction coding (ECC)}, which \mpvi{operates} entirely
within a DRAM chip to improve factory yield. The on-die ECC function and its
effects on DRAM reliability are considered trade secrets, so only the
manufacturer knows \cro{precisely how} on-die ECC alters the externally-visible
reliability characteristics. Consequently, on-die ECC obstructs third-party DRAM
customers \cro{(e.g., test engineers, \mpii{experimental researchers}), who
typically design, test, and validate systems based on these characteristics.}

\cro{To give third parties insight into precisely how on-die ECC transforms DRAM
error patterns during error correction,} we introduce
\underline{B}it-\underline{E}xact \underline{E}CC \underline{R}ecovery (BEER), a
new methodology for determining the full \mpf{DRAM} on-die ECC function (i.e.,
its parity-check matrix) without hardware tools, \mpi{prerequisite knowledge
about the DRAM chip or on-die ECC mechanism, or access to ECC metadata (e.g.,
error syndromes, parity information).} BEER exploits the key insight that
non-intrusively inducing data-retention errors with carefully-crafted test
patterns reveals behavior that is unique to a specific ECC function.

We use BEER to identify the ECC functions of 80 real LPDDR4 DRAM chips with
on-die ECC from three major DRAM manufacturers. We evaluate BEER's correctness
in simulation and performance on a real system to show that BEER is effective
\mpi{and practical across a wide range of on-die ECC functions.} To demonstrate
BEER's value, we propose and discuss several ways that third parties can use
BEER to improve their design and testing practices. As a concrete example, we
introduce and evaluate BEEP, the first error profiling methodology that uses the
known on-die ECC function to recover the number and bit-exact locations of
unobservable raw bit errors responsible for observable post-correction errors.

%% file: 1_introduction.tex
\section{Introduction}
\label{sec:intro}

Dynamic random access memory (DRAM) is the predominant choice for system main
memory across a wide variety of computing platforms due to its favorable
cost-per-bit relative to other memory technologies. DRAM manufacturers maintain
a competitive advantage by improving raw storage densities across device
generations. Unfortunately, \mpi{these improvements largely rely on process
technology scaling, which} causes serious reliability issues that reduce factory
yield. DRAM manufacturers traditionally mitigate yield loss using
post-manufacturing repair techniques such as row/column
sparing~\cite{horiguchi2011nanoscale}. However, continued \mpi{technology
scaling} in modern DRAM chips requires stronger error-mitigation mechanisms to
remain viable because of random single-bit errors that are increasingly frequent
at smaller process technology \jkx{nodes~\cite{micron2017whitepaper, kang2014co,
nair2016xed, gong2017dram, kwon2014understanding, meza2015revisiting, oh20153,
kim2007low, son2015cidra, liu2013experimental, mutlu2014research,
mutlu2013memory}}. Therefore, DRAM manufacturers \mpi{have begun to} use
\emph{on-die error correction \mpi{coding}} \mpi{(\emph{on-die ECC}), which
silently corrects} single-bit errors entirely within the DRAM
chip~\cite{micron2017whitepaper, kang2014co, gong2017dram, nair2016xed,
patel2019understanding}. On-die ECC is \emph{completely invisible} outside of
the DRAM chip, \cro{so} ECC metadata (i.e., parity-check bits, error syndromes)
\mpi{that is used to correct errors} is hidden from the rest of the system.

Prior works~\cite{im2016im, nair2016xed, micron2017whitepaper,
micron2019whitepaper, oh20153, kwak2017a, kwon2017an, patel2019understanding}
indicate that \mpf{existing} on-die ECC codes are 64- or 128-bit single-error
correction (SEC) Hamming codes~\cite{hamming1950error}. However, each DRAM
manufacturer considers their on-die ECC mechanism's design and implementation to
be highly proprietary and ensures not to reveal its details in any public
documentation, including DRAM standards~\cite{jedec2012ddr4, jedec2014lpddr4},
\mpiii{DRAM} datasheets~\mpii{\cite{issi2020lpddr4, samsung2018mobile,
micron2018mobile, hynix2015366ball}}, \jkx{publications~\cite{oh20153,
kwak2017a, kwon2017an, kang2014co}}, and industry
whitepapers~\cite{micron2017whitepaper, micron2019whitepaper}.

Because the unknown on-die ECC function is encapsulated within the DRAM chip,
\mpi{it obfuscates \emph{raw bit} errors (i.e., \emph{pre-correction}
errors)\footnote{\mpi{We use the term ``error'' to refer to \emph{any} bit-flip
event, whether observed (e.g., uncorrectable bit-flips) or unobserved (e.g.,
corrected by ECC).}} in an ECC-function-specific manner. Therefore, the
locations of \mpii{software-visible} \emph{uncorrectable} errors (i.e.,
\emph{post-correction} errors) often no longer match those of the pre-correction
errors} that were caused by physical DRAM error mechanisms. While this behavior
appears desirable from a black-box perspective, it poses serious problems for
third-party DRAM customers who study, test and validate, and/or design systems
based on the reliability characteristics of the DRAM chips that they buy and
use. Section~\ref{subsec:impls_for_third_parties} describes these customers and
the problems they face in detail, including, but not limited to, three important
groups: (1) system designers who need to ensure that supplementary
error-mitigation mechanisms (e.g., rank-level ECC within the DRAM controller)
are carefully designed to cooperate with the on-die ECC
function~\cite{son2015cidra, nair2016xed, gong2018duo}, (2) large-scale
industries \mpii{(e.g., computing system providers such as
Microsoft~\cite{field2015microsoft}, HP~\cite{hp2011whitepaper}, and
Intel~\cite{intel2020platform},} DRAM module
manufacturers~\cite{kingston2012whitepaper, smart2017smart, adata2017adata}) or
government entities (e.g., national labs~\cite{nasa2016nasa,
sandia2020fabrication}) who must understand DRAM reliability characteristics
when validating DRAM chips they buy and use, and (3) researchers who need
\mpf{full visibility into physical device characteristics to study and model}
DRAM \jkx{reliability~\cite{hamamoto1995well, hamamoto1998retention,
jin2004modeling, weis2015retention, edri2016silicon, yaney1987meta,
kim2015avert, patel2019understanding, kim2020revisiting, liu2013experimental,
hassan2017softmc, kim2018dram, kim2018solar, kim2019d, khan2014efficacy,
khan2016case, khan2016parbor, khan2017detecting, chang2016understanding,
chang2017understanding, patel2017reaper}}.

\begin{figure}[h]
    \centering
    \includegraphics[width=\linewidth]{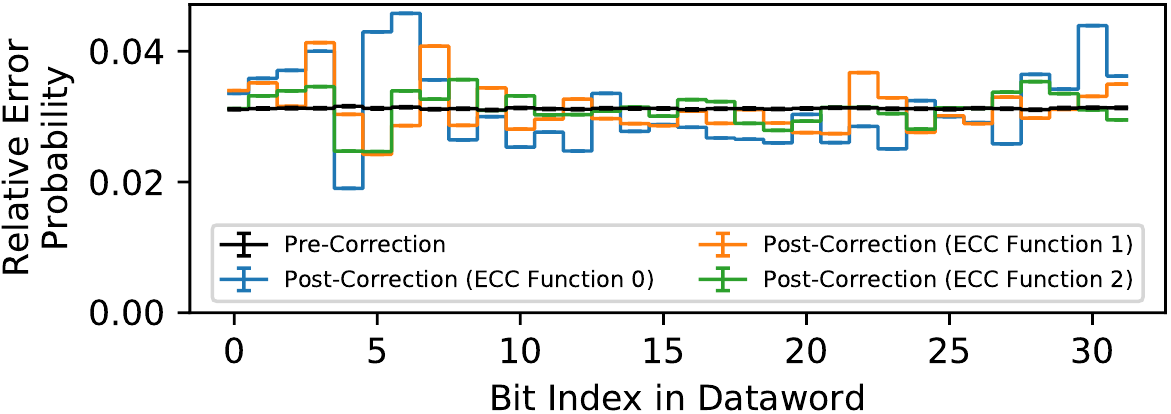}
    \caption{Relative error probabilities in different bit positions for
    different ECC functions with uniform-randomly distributed
    \mpii{pre-correction (i.e., raw)} bit errors.}
    \label{fig:motivation}
\end{figure}

For each of these third parties, merely knowing or reverse-engineering the type
of ECC code (e.g., $n$-bit Hamming code) based on existing
industry~\cite{oh20153, kwak2017a, kwon2017an, micron2017whitepaper,
micron2019whitepaper, im2016im} and academic~\cite{nair2016xed,
patel2019understanding} publications is not enough to \mpi{determine} exactly
how the ECC mechanism obfuscates specific error patterns. This is because an ECC
code of a given type can have many different implementations based on how its
ECC function (i.e., its parity-check matrix) is designed, and different designs
lead to different reliability characteristics. For example,
Figure~\ref{fig:motivation} shows the relative probability of observing errors
in different bit positions for three different ECC codes of the same type (i.e.,
\mpvi{single-error correction} Hamming code \mpii{with 32 data bits and 6
parity-check bits}) but that use different ECC functions. We obtain this data by
simulating $10^9$ ECC words using the EINSim
simulator~\cite{patel2019understanding, eccsimgithub} and show medians and 95\%
confidence intervals calculated via statistical
bootstrapping~\cite{efron1992bootstrap} over 1000 samples. \mpi{We simulate a
\texttt{0xFF} test pattern\footnote{Other patterns show similar behavior,
including \texttt{RANDOM} data.} with uniform-random pre-correction errors at} a
raw bit error rate of $10^{-4}$ (e.g., \mps{as often seen in} experimental
studies\hht{~\cite{hamamoto1998retention, kang2014co, liu2013experimental,
lee2015adaptive, shirley2014copula, chang2017understanding, hassan2017softmc,
chang2016understanding, patel2017reaper}}).

\mpf{The data demonstrates that ECC codes of the same type can have vastly
different post-correction error characteristics. This is because each ECC
mechanism acts differently when faced with more errors than it can correct
(i.e., uncorrectable errors), causing it to mistakenly perform
ECC-function-specific ``corrections'' to bits that did not experience errors
(i.e., \mpvi{\emph{miscorrections}}, which Section~\ref{subsec:ondie_ecc} expands upon).
Therefore, a researcher or engineer who studies two DRAM chips that use the same
type of ECC code but different ECC functions may find that the chips'
\mpiii{software-visible} reliability characteristics are quite different even if
the \mpiii{physical} DRAM cells' reliability characteristics are identical. On
the other hand, if we know the full ECC function (i.e., its parity-check
matrix), we can calculate exactly which pre-correction error pattern(s) result
in a set of observed errors. Figure~\ref{fig:motivation} is a result of
aggregating such calculations across \mpii{$10^9$} error
patterns\footnote{\mpvi{Capturing approximately $10^9$ of the $2^{38} \approx
2.7\times 10^{11}$ unique patterns.}}, and Section~\ref{subsec:beep}
demonstrates how we can use the ECC function to infer pre-correction
error counts and locations using \mps{only} observed post-correction errors.}

\mpf{Knowing the precise transformation between pre- and post-correction errors}
benefits all of the \mpi{aforementioned} third-party use cases \mpf{because it
provides system designers, test engineers, and researchers \mpi{with} a way to
isolate the error characteristics of the memory itself from the effects of a
particular ECC function.} Section~\ref{subsec:impls_for_third_parties} provides
\mpf{several example use cases and describes the benefits of knowing the ECC
function in detail. \mpi{While} specialized, possibly intrusive methods (e.g.,
chip teardown~\mpvi{\cite{james2010silicon, torrance2009state}}, advanced
imaging techniques~\cite{ho2003method, torrance2009state}) can theoretically
extract the ECC function, such techniques are typically inaccessible to or
infeasible for many third-party users.}

To enable third parties to \mpf{reconstruct pre-correction DRAM reliability
characteristics}, \textbf{our goal} is to develop a methodology that can
reliably and accurately determine the \mpi{full on-die ECC function without
requiring hardware tools, prerequisite knowledge about the DRAM chip or on-die
ECC mechanism, or access to ECC metadata (e.g., error syndromes, parity
information)}. To this end, we develop \underline{B}it-\underline{E}xact
\underline{E}CC \underline{R}ecovery (BEER), a new methodology for determining
\cro{a DRAM chip's full on-die ECC function simply by studying the
software-visible post-correction error patterns that it generates}. Thus, BEER
requires no hardware support, hardware intrusion, or access to internal ECC
metadata (e.g., error syndromes, parity information). BEER exploits the key
insight that forcing the ECC function to act upon carefully-crafted
uncorrectable error patterns reveals ECC-function-specific behavior that
disambiguates different ECC functions. BEER comprises three key steps: (1)
\cro{deliberately} inducing uncorrectable data-retention errors by \cro{pausing
DRAM refresh} while using carefully-crafted test patterns to control \mpf{the
errors'} bit-locations, which is done by leveraging data-retention errors'
intrinsic \mpf{data-pattern asymmetry} (discussed in
Section~\ref{subsec:dram_errors}), (2) enumerating the bit positions where the
ECC mechanism causes miscorrections, and (3) using a SAT solver~\cite{de2008z3}
to solve for the unique parity-check matrix \mpf{that causes} the observed set
of miscorrections.

We experimentally apply BEER to 80 real LPDDR4 DRAM chips with on-die ECC from
three major DRAM manufacturers to determine \mpii{the chips'} on-die ECC
functions. We describe the experimental steps required to apply BEER to any DRAM
chip with on-die ECC and show that BEER tolerates observed experimental noise.
We show that different manufacturers appear to use different on-die ECC
functions while chips from the same manufacturer and model number appear to use
the same on-die ECC function
\mpi{(Section~\ref{subsubsec:testing_charged_patterns})}. \cro{Unfortunately},
our experimental studies with real DRAM chips have two limitations against
further validation: (1) because the on-die ECC function is considered trade
secret for each manufacturer, we are unable to obtain a groundtruth to compare
BEER's results against, even when considering non-disclosure agreements with
DRAM manufacturers and (2) we are unable to publish the final ECC functions that
we uncover using BEER for confidentiality reasons (discussed in
Section~\ref{subsec:motivation_secrecy}).

To overcome the limitations of experimental studies with real DRAM chips, we
rigorously evaluate BEER's correctness in simulation
(Section~\ref{sec:beer1eval}). We show that BEER correctly recovers the on-die
ECC function for 115300 single-error correction Hamming codes\footnote{This
irregular number arises from evaluating a \mpii{different} number of ECC
functions for different code lengths because longer codes require exponentially
more simulation time (discussed in
Section~\ref{subsec:beer1eval_correctness}).}, which are representative of
on-die ECC, with ECC word lengths ranging from 4 to 247 bits. We evaluate our
BEER implementation's runtime and memory consumption using a real system to
demonstrate that \mpii{BEER is practical and} the SAT problem that BEER requires
is realistically solvable.

\cro{To demonstrate how BEER \cro{is} useful in practice, we propose and discuss
several ways that third parties can leverage the ECC function that BEER reveals
to \mpi{more effectively design, study, and test systems that use DRAM chips
with on-die ECC} (Section~\ref{section:usecases}). As a concrete example, we
introduce and} evaluate \underline{B}it-\underline{E}xact \underline{E}rror
\underline{P}rofiling (BEEP), a new DRAM data-retention error profiling
methodology that reconstructs pre-correction error counts and locations purely
from observed post-correction errors. Using the ECC function revealed by BEER,
BEEP infers precisely which \emph{unobservable} raw bit errors correspond to
\emph{observed} post-correction errors at a given set of testing conditions. We
show that BEEP enables characterizing pre-correction errors across a wide range
of ECC functions, \mpi{ECC word} lengths, error patterns, and error rates.
\mpii{We publicly release our tools as open-source \mpii{software}: (1) a new
tool~\cite{beergithub} for applying BEER to experimental data from real DRAM
chips and (2) enhancements to EINSim~\cite{eccsimgithub} for evaluating BEER
\mpii{and BEEP} in simulation.}

This paper makes the following key contributions:
\begin{enumerate}
    \item We provide Bit-Exact ECC Recovery (BEER), the first methodology that
    determines the full DRAM on-die ECC function (i.e., its parity-check matrix)
    without \mpvi{requiring hardware tools, prerequisite knowledge about the
    DRAM chip or on-die ECC mechanism}, or access to ECC metadata (e.g., error
    syndromes, parity information).

    \item We experimentally apply BEER to 80 real LPDDR4 DRAM chips with unknown
    on-die ECC mechanisms from three major DRAM manufacturers to determine their
    on-die ECC functions. We show that BEER is robust to observed experimental
    noise and that \mpiii{DRAM chips from} different manufacturers \mpf{appear
    to} use different on-die ECC functions while chips from the same
    manufacturer and model number \mpf{appear to} use the same function.

    \item We evaluate BEER's correctness in simulation and show that BEER
    correctly identifies the \mpi{on-die ECC function for 115300 representative
    on-die ECC codes with ECC word lengths ranging from 4 to 247 bits.}

    \item We analytically evaluate BEER's experimental runtime and use a real
    system to measure the SAT solver's performance and memory usage
    characteristics (e.g., negligible for short codes, \mpii{median of 57.1
    hours and 6.3 GiB memory for representative 128-bit codes, and} up to 62
    hours and 11.4 GiB memory for 247-bit codes) to show that BEER is practical.
    
    \item \mpf{We propose and evaluate Bit-Exact Error Profiling (BEEP), a new
    DRAM data-retention error profiling methodology that uses a known ECC
    function (e.g., via BEER) to infer pre-correction error counts and
    locations. We show that BEEP enables characterizing the bit-exact
    pre-correction error locations across} different ECC functions, codeword
    lengths, error patterns, and error rates.
    
    \item We open-source the software tools we develop for (1) applying BEER to
    experimental data from real DRAM chips~\cite{beergithub} and (2) \mpi{evaluating
    BEER and BEEP in simulation~\cite{eccsimgithub}.}
\end{enumerate}

%% file: 2_motivation.tex
\section{Challenges of Unknown On-Die ECCs}
\label{sec:motivation}

This section discusses why on-die ECC is considered proprietary, how its secrecy
causes difficulties for third-party consumers, and how \mpi{the BEER methodology
helps overcome these difficulties \mpii{by} identifying} the \mpii{full} on-die ECC
function.

\subsection{Secrecy Concerning On-Die ECC}
\label{subsec:motivation_secrecy}

On-die ECC silently mitigates increasing single-bit errors that reduce factory
\jkx{yield~\cite{micron2017whitepaper, kang2014co, nair2016xed, gong2017dram,
kwon2014understanding, meza2015revisiting, oh20153, kim2007low, son2015cidra,
liu2013experimental, mutlu2014research, mutlu2013memory}}. Because on-die ECC is
invisible to the external DRAM chip interface, older DRAM
standards~\cite{jedec2014lpddr4, jedec2012ddr4} place no restrictions \mpvi{on
the on-die ECC mechanism} while newer standards~\cite{jedec2020ddr5} specify
only a high-level description for on-die ECC to support new (albeit limited)
DDR5 features, e.g., on-die ECC scrubbing. In particular, there are no
restrictions on the \mpvi{design or implementation of the on-die ECC function
itself.}

This means that knowing \mpi{an on-die ECC mechanism's details could reveal
information about its} manufacturer's factory yield rates, which are highly
proprietary~\cite{cost1997yield, childers2015achieving} due to their direct
connection with business interests, potential legal concerns, and
competitiveness in a USD 45+ billion DRAM market~\cite{qy2019global,
verified2019global}. Therefore, manufacturers consider their on-die ECC designs
and implementations to be trade secrets that they are unwilling to disclose.
\revmp{In our experience, DRAM manufacturers will not reveal on-die ECC details
under confidentiality agreements, even for large-scale industry board vendors
for whom knowing the details stands to be mutually
beneficial.}\footnote{\revmp{Even if such agreements were possible, industry
teams and academics without major business relations with DRAM manufacturers
(i.e., an overwhelming majority of the potentially interested scientists and
engineers) will likely be unable to secure disclosure.}}

This raises two challenges for our experiments with real DRAM chips: (1) we do
not have access to ``groundtruth'' ECC functions to validate BEER's results
against and (2) we cannot publish the final ECC functions that we determine
using BEER for confidentiality reasons based on our relationships with the DRAM
manufacturers. However, this does not prevent third-party consumers from
applying BEER to their own devices, and we hope that our work encourages DRAM
manufacturers to be more open with their designs going forward.\footnote{While
full disclosure would be ideal, a more realistic scenario could be more flexible
on-die ECC confidentiality agreements. \mpi{As recent
work~\cite{frigo2020trrespass} shows, security or protection by obscurity is
likely a poor strategy in practice.}}

\subsection{On-Die ECC's Impact on Third Parties}
\label{subsec:impls_for_third_parties}

On-die ECC alters a DRAM chip's \mpii{software-visible} reliability
characteristics so that \mpi{they are no longer determined solely by how errors
physically occur within \mpi{the DRAM chip}. Figure~\ref{fig:motivation}
illustrates this by showing how \mpi{using} different \mpi{on-die} ECC functions
changes how the \emph{same} underlying DRAM errors appear to the end user.
Instead of following the pre-correction error distribution (i.e., uniform-random
errors), the post-correction errors exhibit ECC-function-specific shapes that
\mpiii{are difficult to predict} without knowing precisely which ECC function
\mpiii{is} used in each case. This means that two commodity DRAM chips with
different on-die ECC functions may show similar or different reliability
characteristics irrespective of how the underlying DRAM technology and error
mechanisms behave. Therefore, the \mpii{physical} error mechanisms' behavior
alone can no longer explain a DRAM chip's  post-correction error
characteristics.}

\mpi{Unfortunately, this poses a serious problem for third-party DRAM consumers
(e.g., system designers, testers, and researchers), who can no longer
\mpii{accurately} understand a DRAM chip's reliability characteristics by
studying its \mpii{software-visible} errors. This lack of understanding prevents
third parties from \mpii{both (1)} making informed design decisions, e.g., when
building \mpii{memory-controller based} error-mitigation mechanisms to
complement on-die ECC and \mpii{(2)} developing new ideas \mpii{that rely on} on
leveraging predictable aspects of a DRAM chip' reliability characteristics,
e.g., physical error mechanisms that are fundamental to all DRAM technology.}
As error rates worsen with continued technology
\jkx{scaling~\cite{micron2017whitepaper, kang2014co, nair2016xed, gong2017dram,
kwon2014understanding, meza2015revisiting, oh20153, kim2007low,
kim2020revisiting, kim2014flipping, mutlu2013memory, mutlu2014research}},
manufacturers will likely resort to stronger codes that further distort the
post-correction reliability characteristics. The remainder of this section
describes three key ways in which an unknown on-die ECC function hinders
third-parties, and determining the function helps mitigate the problem.

\noindent
\textbf{Designing \mpt{High-Reliability} Systems.} System designers often seek
to improve memory reliability beyond that which the DRAM provides alone (e.g.,
\mpi{by including rank-level ECC within} the memory controllers of server-class
machines or ECC within on-chip caches). In particular, rank-level ECCs are
carefully designed to mitigate common DRAM failure
modes~\cite{chen2018configurable} (e.g., chip failure~\cite{nair2016xed}, burst
errors~\cite{maiz2003characterization, dell1997white}) \mpi{in order to correct
as many errors as possible. However, designing for key failure modes requires
knowing a DRAM chip's reliability characteristics, including the effects of any
underlying ECC function (e.g., on-die ECC)~\cite{son2015cidra, gong2018duo}. For
example, Son et al.~\cite{son2015cidra} show that if on-die ECC suffers an
uncorrectable error and mistakenly ``corrects'' a non-erroneous bit (i.e.,
introduces a \emph{miscorrection}), the stronger rank-level ECC may no longer be
able to even detect what would otherwise be a detectable (possibly correctable)
error. To prevent this scenario, both levels of ECC must be carefully
co-designed to complement each others' weaknesses.} In general, high-reliability
systems can be more effectively built around DRAM chips with on-die ECC if its
ECC function and its effects on typical DRAM failure modes are known.

\noindent
\textbf{\mpf{Testing}, Validation, and Quality Assurance.} \mpf{Large-scale
\mpi{computing system providers} \mpf{(e.g.,
Microsoft~\cite{field2015microsoft}, HP~\cite{hp2011whitepaper},
Intel~\cite{intel2020platform}), DRAM module
manufacturers~\cite{kingston2012whitepaper, smart2017smart, adata2017adata}, and
government entities (e.g., national labs~\cite{sandia2020fabrication,
nasa2016nasa}) typically} perform extensive third-party testing of the DRAM
chips they purchase in order to ensure that the chips meet internal
performance/energy/reliability targets. These tests validate} that DRAM chips
operate as expected \mpi{and that there are \mpii{well-understood, convincing}
root-causes (e.g., fundamental DRAM error mechanisms) for any observed errors}.
Unfortunately, on-die ECC interferes with two key components of such testing.
First, it obfuscates the number and bit-exact locations \mpi{of pre-correction
errors, so diagnosing the root cause for any observed \mpii{error} becomes
challenging.} Second, on-die ECC encodes all written data into \mpi{ECC
codewords, so the values written into the physical cells likely do not match the
values observed at the DRAM chip interface. The encoding process defeats
carefully-constructed test patterns that target specific circuit-level phenomena
(e.g., exacerbating interference between bitlines~\cite{adams2002high,
mrozek2019multi, khan2016parbor}) because the encoded data may no longer have
the intended effect. Unfortunately, constructing such patterns is crucial for
efficient testing since it minimizes the testing time required to achieve high
error coverage}~\cite{horiguchi2011nanoscale, adams2002high}. In both cases, the
full on-die ECC function determined by BEER describes exactly how on-die ECC
transforms pre-correction error patterns into post-correction ones. This enables
users to infer pre-correction error locations (demonstrated in
Section~\ref{subsec:beep}) and design test patterns that result in codewords
with desired properties (discussed in Section~\ref{subsection:use_cases}).

\noindent
\textbf{Scientific \mpvi{Error-Characterization} Studies.} 
Scientific error-characterization studies explore physical DRAM error mechanisms
(e.g., data retention~\hht{\cite{hamamoto1995well, hamamoto1998retention,
shirley2014copula, jung2015omitting, weis2015thermal, weis2015retention,
jung2014optimized, khan2016parbor, khan2016case, khan2014efficacy,
liu2013experimental, patel2017reaper, hassan2017softmc, khan2017detecting}},
reduced access-latency~\cite{chang2016understanding, lee2015adaptive,
lee2017design, kim2018solar, kim2019d, kim2018dram, gao2019computedram,
chang2017understanding, chandrasekar2014exploiting}, circuit
disturbance~\hht{\cite{kim2014flipping, khan2016parbor, park2016statistical,
park2016experiments, kim2020revisiting, frigo2020trrespass, khan2017detecting}})
by deliberately exacerbating the error mechanism and analyzing the resulting
errors' statistical properties (e.g., frequency, spatial distribution). These
studies help build error models~\cite{yaney1987meta, hamamoto1998retention,
shirley2014copula, edri2016silicon, lee2017design, chang2017understanding,
kim2018solar, koppula2019eden}, leading to new DRAM designs and operating points
that improve upon the state-of-the-art. Unfortunately, on-die ECC complicates
error analysis and modeling by (1) \mpi{obscuring the physical pre-correction
errors that are the object of study} and (2) \mpi{preventing direct access to
parity-check bits, thereby precluding comprehensive testing of all DRAM cells in
a given chip.} Although prior work~\cite{patel2019understanding} enables
inferring high-level statistical characteristics of the pre-correction errors,
it does not provide a precise mapping between pre-correction and post-correction
errors, which is only possible knowing the full ECC function. Knowing the full
ECC function\mpi{, via our new BEER methodology,} enables recovering the
bit-exact locations of pre-correction errors throughout the entire ECC word
(\mpi{as we demonstrate} in Section~\ref{subsec:beep}) so that
error-characterization studies \mpi{can separate the effects of DRAM error
mechanisms from those of on-die ECC}. Section~\ref{section:usecases} provides a
detailed discussion of several key characterization studies that BEER enables.

%% file: 3_background.tex
\section{Background}

This section provides a basic overview of DRAM, coding theory, and
satisfiability (SAT) solvers as pertinent to this manuscript. For further
detail, we refer the reader to comprehensive texts on DRAM design and
operation~\jkz{\cite{keeth2007dram, jacob2010memory, iniewski2011nano,
itoh2013vlsi, hassan2019crow, seshadri2017ambit, seshadri2019dram,
chang2017understanding, chang2016understanding, lee2015adaptive, zhang2014half,
ipek2008self, chang2014improving, lee2015decoupled, lee2015simultaneous,
chang2016low, luo2020clr, seshadri2013rowclone}}, coding
theory~\cite{macwilliams1977theory, moon2005error, roth2006introduction,
huffman2010fundamentals, richardson2008modern, clark2013error, costello2004ecc},
and SAT solvers~\cite{de2008z3, bjorner2015nuz, cimatti2010satisfiability,
dillig2012minimum}.

\subsection{DRAM Cells and Data Storage}

A DRAM chip stores each data bit in its own \emph{storage cell} using the
\jkz{charge} level of a \emph{storage capacitor}. Because the capacitor is
susceptible to charge leakage~\jkz{\cite{hamamoto1995well,
venkatesan2006retention, patel2017reaper, liu2012raidr, cojocar19exploiting,
kraft2018improving, kim2014flipping, liu2013experimental,
patel2019understanding}}, the stored value may eventually degrade to the point
of data loss, resulting in a \emph{data-retention error}. During normal DRAM
operation, a \emph{refresh} operation restores the data value stored in each
cell every \emph{refresh window} (\emph{$t_{REFw}$}), e.g., 32ms or
64ms~\jkz{\cite{jedec2014lpddr4, jedec2012ddr4, jedec2008ddr3,
liu2013experimental, patel2017reaper, liu2012raidr}}, to prevent data-retention
errors.

Depending on a given chip's circuit design, each cell may store data using one
of two encoding conventions: a \mpvi{\emph{true-cell}} encodes data `1' as \cro{a
fully-charged storage capacitor (i.e., the \texttt{CHARGED} state), and an
\mpvi{\emph{anti-cell}} encodes data `1' as a fully-discharged capacitor (i.e., the
\texttt{DISCHARGED} state). \mpiii{Although a cell's encoding
scheme} is transparent to the rest of the system during normal operation, it
becomes evident in the presence of data-retention errors because DRAM cells
typically decay only from their \texttt{CHARGED} to their \texttt{DISCHARGED}
state as shown experimentally by prior work~\cite{cojocar19exploiting,
kraft2018improving, kim2014flipping, liu2013experimental,
patel2019understanding, patel2017reaper, liu2012raidr}.}

\subsection{\mpt{Studying DRAM Errors}}
\label{subsec:dram_errors}

Deliberately inducing DRAM errors (e.g., by violating default \mpii{timing
parameters}) reveals detailed information about a DRAM chip's internal design
through the resulting errors' statistical characteristics. Prior works use
custom memory testing platforms (e.g., FPGA-based~\cite{hassan2017softmc}) and
commodity CPUs~\cite{amd_opteron_mcconfig, intel_4thgen_mcconfig} (e.g., by
changing CPU configuration registers via the BIOS~\cite{intel_xmp}) to study
a variety of DRAM error mechanisms, including
data-retention~\jkz{\cite{patel2017reaper, liu2012raidr, cojocar19exploiting,
kraft2018improving, kim2014flipping, liu2013experimental,
patel2019understanding}}, circuit timing violations~\jkz{\cite{kim2018solar,
lee2017design, chang2016understanding, lee2015adaptive, kim2019d,
kim2018dram}}, and RowHammer~\jkz{\cite{kim2014flipping, mutlu2017rowhammer,
mutlu2019rowhammer, kim2020revisiting, park2016statistical,
park2016experiments}}. Our work focuses on data-retention errors because they
exhibit well-studied properties that are \jkz{helpful for our purposes:}
\begin{enumerate}
    \item \jk{They are easily} induced and controlled by manipulating the
    refresh window ($t_{REFw}$) and ambient temperature.
	\item \revmp{They are repeatable~\cite{sutar2016d, patel2017reaper}
	\jkz{and their spatial distribution is uniform
	random}~\cite{baek2014refresh, hamamoto1998retention, kim2018dram,
	shirley2014copula, patel2019understanding}.}    
    \item They fail unidirectionally from the \texttt{CHARGED} state to
	the \texttt{DISCHARGED} \jkz{state}~\cite{cojocar19exploiting,
  kraft2018improving, kim2014flipping, liu2013experimental,
  patel2019understanding, patel2017reaper, liu2012raidr}.
\end{enumerate}

\noindent
\revmpf{\textbf{\mpi{Off-DRAM-Chip Errors}.}}
\revmpf{Software-visible memory errors often occur due to failures in components
outside the DRAM chip \mpii{(e.g., sockets, buses)}~\cite{meza2015revisiting}. However,
our work focuses on errors that occur \emph{within} a DRAM chip, which are a
serious and growing concern at modern technology node
\jkx{sizes~\cite{micron2017whitepaper, kang2014co, nair2016xed, gong2017dram,
kwon2014understanding, meza2015revisiting, oh20153, kim2007low, son2015cidra,
mutlu2013memory, mutlu2014research}}.  These errors are the primary motivation
for on-die ECC, which attempts to correct them before they are ever observed
outside the DRAM chip.}

\subsection{On-Die ECC and Hamming Codes}
\label{subsec:ondie_ecc}

As manufacturers continue to increase DRAM storage density, unwanted
single-bit errors \mpt{appear more frequently}~\jkz{\cite{lee2013tiered,
nair2013archshield, nair2016xed, micron2017whitepaper, kang2014co,
hong2010memory, gu2003challenges, meza2015revisiting, sridharan2012study,
sridharan2015memory, luo2014characterizing, schroeder2009dram,
patel2019understanding, gong2017dram, kwon2014understanding, oh20153,
kim2007low}} and \mpt{reduce} factory yield. To combat these errors,
manufacturers \jkz{use} on-die ECC~\jkz{\cite{nair2013archshield, nair2016xed,
micron2017whitepaper, kang2014co, oh20153, gong2017dram,
patel2019understanding}}, which is an error-correction code implemented
directly in the DRAM chip.

\begin{figure}[h]
    \centering
    \includegraphics[width=\linewidth]{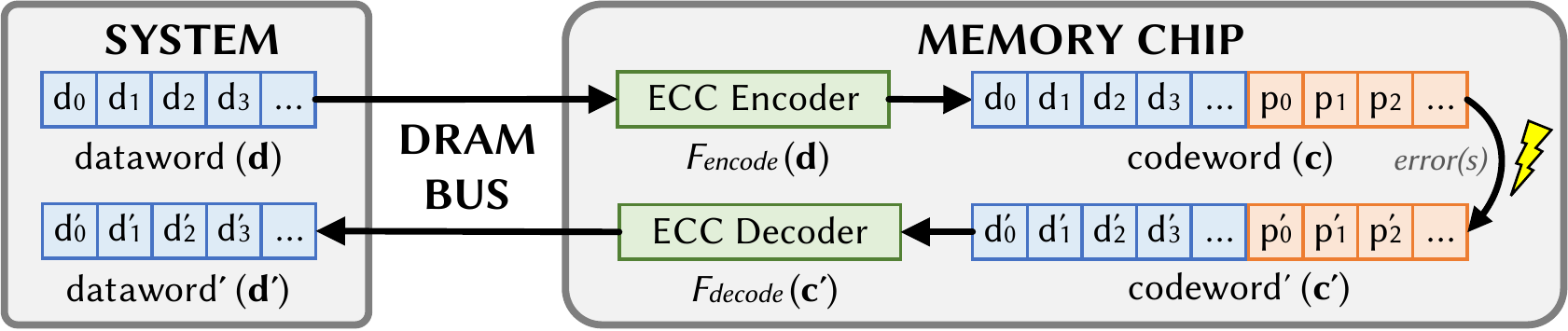}
    \caption{\mpii{Interfacing a memory chip that uses} on-die ECC.}
    \label{fig:ondieecc}
\end{figure}

Figure~\ref{fig:ondieecc} shows how a system might interface with a memory chip
that uses on-die ECC. The system writes $k$-bit \emph{datawords} ($\mathbf{d}$)
to the chip, which internally maintains an expanded $n$-bit representation of
the data called a \emph{codeword} ($\mathbf{c}$), \jkz{created by the ECC
encoding of d}. The stored codeword may experience errors, resulting in a
potentially erroneous codeword ($\mathbf{c'}$). If more errors occur than ECC
can correct, e.g., two errors in a single-error correction (SEC) code, the final
dataword read out \jkz{after ECC decoding} ($\mathbf{d'}$) may also contain
errors. The encoding and decoding functions are labeled $F_{encode}$ and
$F_{decode}$.

For all linear codes (e.g., SEC Hamming
codes~\cite{hamming1950error}), $F_{encode}$ and $F_{decode}$ can be
represented using matrix transformations. \jk{As a demonstrative example
throughout} this paper, we use the (7, 4, 3) Hamming
\jkz{code~\cite{hamming1950error}} shown in Equation~\ref{eqn:hg}. $F_{encode}$
represents a generator matrix $\mathbf{G}$ such that the codeword $\mathbf{c}$
is computed from the dataword
$\mathbf{d}$ as $\mathbf{c} = \mathbf{G} \cdot \mathbf{d}$.
\vspace{-0.25\baselineskip}
\begin{equation}
\vspace{-0.5\baselineskip}
\arraycolsep=0.9pt
\def\arraystretch{0.5}
\small
F_{encode}=\mathbf{G^T}=\left[
  \begin{array}{cccc|ccc}
    1 & 0 & 0 & 0 & 1 & 1 & 1 \\
    0 & 1 & 0 & 0 & 1 & 1 & 0 \\
    0 & 0 & 1 & 0 & 1 & 0 & 1 \\
    0 & 0 & 0 & 1 & 0 & 1 & 1
  \end{array}
\right]
\quad
F_{decode}=\mathbf{H}=\left[
  \begin{array}{cccc|ccc}
    1 & 1 & 1 & 0 & 1 & 0 & 0 \\
    1 & 1 & 0 & 1 & 0 & 1 & 0 \\
    1 & 0 & 1 & 1 & 0 & 0 & 1 
  \end{array}
\right]
\label{eqn:hg}
\end{equation}

\noindent
\textbf{Decoding.}
The most common decoding algorithm is known as \emph{syndrome decoding}, which
simply computes an \emph{error syndrome} $\mathbf{s} = \mathbf{H} \cdot
\mathbf{c'}$ that describes if and where an error exists:
\begin{itemize}
\item $\mathbf{s}=\mathbf{0}$: no error detected.
\item $\mathbf{s}\neq\mathbf{0}$: error detected, and $\mathbf{s}$ describes its bit-exact location.
\end{itemize}
\noindent
Note that the error syndrome computation is \emph{unaware} of the true error
count; it blindly computes the error syndrome(s) assuming a low probability of
uncorrectable errors. If, however, \jk{an} uncorrectable error is present (e.g.,
deliberately induced during testing), one of three possibilities may occur:
\begin{itemize}
\item \emph{Silent data corruption:} syndrome is zero; \mpt{no error}.
\item \emph{Partial correction:} syndrome points to one of the errors.
\item \emph{Miscorrection:} syndrome points to a non-erroneous bit.
\end{itemize}
\noindent
\mpi{When a nonzero error syndrome occurs, the ECC decoding logic simply flips
the bit pointed to by the error syndrome, potentially \emph{exacerbating} the
overall number of errors.}

\noindent
\textbf{Design Space.} \jkz{Each manufacturer can freely select $F_{encode}$ and
$F_{decode}$ functions, whose implementations can help to meet a set of design
constraints (e.g., circuit area, reliability, power consumption). The space of
functions that a designer can choose from is quantified by the number of
arrangements of columns of $\mathbf{H}$. This means that for an $n$-bit code
with $k$ data bits, there are ${2^{n-k} - 1 \choose n}$ possible ECC functions.
Section~\ref{subsubsec:determining_func_from_mcp} formalizes this space of
possible functions in the context of our work.} 

\subsection{Boolean Satisfiability (SAT) Solvers}
\label{subsec:sat_solver}

\mpo{Satisfiability (SAT) solvers~\jkz{\cite{de2008z3, bjorner2015nuz,
cimatti2010satisfiability, dillig2012minimum, gomes2008satisfiability,
prasad2005survey}} \mpf{find} possible solutions to logic equation(s)
\jkz{with} one or more unknown Boolean variables.} A SAT solver accepts one or
more such equations as inputs, which effectively act as \emph{constraints} over
the unknown variables.  The SAT solver then attempts to determine a set of
values for the unknown variables such that the equations are satisfied (i.e.,
the constraints are met). \mpo{The SAT solver will return either (1) one (of
possibly many) solutions or (2)} no solution if the Boolean equation is
unsolvable.

%% file: 4_beer.tex
\section{Determining the ECC Function}
\label{sec:determining_ecc_func}

BEER identifies an \mpi{unknown} ECC function by systematically reconstructing
its parity-check matrix based on the error syndromes that the ECC logic
generates while correcting errors. Different ECC functions compute different
error syndromes for a given error pattern, and by \mpi{constructing and
analyzing} carefully-crafted test cases, BEER uniquely identifies which ECC
function a particular implementation uses. This section describes how and why
this process works. Section~\ref{sec:beeri} describes how BEER accomplishes this
in practice for on-die ECC.

\subsection{\mpt{Disambiguating} Linear Block Codes}
\label{subsubsec:systematically_identifying}

DRAM ECCs are linear block codes, e.g., Hamming codes~\cite{hamming1950error}
for on-die ECC~\cite{im2016im, nair2016xed, micron2017whitepaper,
micron2019whitepaper, oh20153, kwak2017a, kwon2017an, patel2019understanding},
BCH~\cite{bose1960class, hocquenghem1959codes} or
Reed-Solomon~\cite{reed1960polynomial} codes for rank-level
ECC~\cite{cojocar19exploiting, kim2016all}, whose encoding and decoding
operations are described by \emph{linear transformations} of their respective
inputs (i.e., $\mathbf{G}$ and $\mathbf{H}$ \mpi{matrices}, respectively).
\cro{We can therefore determine the full ECC function by independently
determining each of its linear components.}

\mpt{We can isolate each linear component of the ECC function} by injecting
errors in each codeword bit position and observing the resulting error
syndromes. For example, an $n$-bit Hamming code's parity-check matrix can be
systematically determined by injecting a single-bit error in each of the $n$ bit
positions: the error syndrome that the ECC decoder computes for each pattern is
exactly equal to the column of the parity-check matrix that corresponds to
\mpi{the} position of the injected error. As an example, Equation~\ref{eqn:syn}
shows how injecting an error at position 2 (i.e., adding error pattern
$\mathbf{e_2}$ to codeword $\mathbf{c}$) extracts the corresponding column of
the parity-check matrix $\mathbf{H}$ in the error syndrome $\mathbf{s}$. \mpi{By
the definition of a block code, $\mathbf{H}\cdot\mathbf{c}=\mathbf{0}$ for all
codewords~\cite{costello1982error, huffman2010fundamentals}, so} $\mathbf{e_2}$
isolates column 2 of $\mathbf{H}$ (i.e., $\mathbf{H}_{\ast,2}$).
\vspace{-0.15\baselineskip}
\begin{equation}
\vspace{-0.25\baselineskip}
\arraycolsep=1.6pt
\def\arraystretch{0.2}
\footnotesize
\mathbf{s}=\mathbf{H}\cdot\mathbf{c'}=\mathbf{H}\cdot(\mathbf{c} + \mathbf{e_2})=\mathbf{H}\cdot\left(\mathbf{c}+\left[
  \begin{array}{c}
    0 \\ 0 \\ 1 \\ 0 \\ 0 \\ 0 \\ 0
  \end{array}
\right]\right)=\mathbf{0} + \mathbf{H}_{\ast,2} = \mathbf{H}_{\ast,2}
\label{eqn:syn}
\end{equation}
Thus, the entire parity-check matrix can be fully determined by testing across
all 1-hot error patterns. \jk{Cojocar et al.~\cite{cojocar19exploiting} use
this approach on DRAM rank-level ECC, injecting errors into codewords on the
DDR bus and reading the resulting error syndromes provided by the memory controller.} 

\subsection{Determining the On-Die ECC Function}
\label{subsubsec:determining_func_from_mcp}

\jk{Unfortunately, systematically determining \mpf{an ECC function} as described
in Section~\ref{subsubsec:systematically_identifying} is not possible with
on-die ECC for two key reasons.} First, \mpf{on-die ECC's parity-check bits
\mpi{cannot be accessed directly}, so we have no easy way to inject an error
within them.} Second, on-die ECC does not signal an error-correction event
\emph{or} report error syndromes (i.e., $\mathbf{s}$). \mpf{Therefore, even if
\mpvi{specialized methods} \mpi{(e.g., chip teardown~\cite{james2010silicon,
torrance2009state}, \mpvi{advanced imaging techniques}~\cite{ho2003method,
torrance2009state})} could inject errors within a DRAM chip package where the
on-die ECC mechanism resides,}\footnote{\mpi{Such methods may reveal the exact
on-die ECC circuitry. However, they are typically inaccessible to or infeasible
for many third-party consumers.}} the error syndromes would remain invisible, so
the approach taken by Cojocar et al.~\cite{cojocar19exploiting} cannot be
applied to on-die ECC. To \mpf{determine the on-die ECC function using the}
approach of Section~\ref{subsubsec:systematically_identifying}, we first
formalize the unknown on-die ECC function and then determine how we can infer
error syndromes within the constraints of the formalized problem.

\subsubsection{Formalizing the Unknown ECC Function}
\label{subsubsec:formalizing_ecc_func}

We assume that on-die ECC uses a systematic encoding, which means that the ECC
function stores data bits unmodified. This is a reasonable assumption for real
hardware since it greatly simplifies data access~\cite{zhang2015vlsi} and is
consistent with our experimental results in
Section~\ref{subsubsec:ecc_word_layout}. Furthermore, \mpi{because the DRAM chip
interface exposes only data bits, the relative ordering of parity-check bits
within the codeword is irrelevant from the system's perspective.}
Mathematically, the different choices of bit positions represent
\emph{equivalent codes} that all have identical error-correction properties and
differ only in their internal
representations~\cite{richardson2008modern,roth2006introduction}, which on-die
ECC does not expose. Therefore, we are free to arbitrarily choose the
\mpi{parity-check bit} positions within the codeword without loss of generality.
If it becomes necessary to identify the exact ordering of bits within the
codeword \mpiii{(e.g., to infer circuit-level implementation details)},
reverse-engineering techniques based on physical DRAM error
mechanisms~\cite{lee2017design, jung2016reverse} can potentially be used.

A systematic encoding and the freedom to choose parity-check bit positions
mean that we can assume that the ECC function is in \emph{standard form},
where we express the parity-check matrix for an $(n, k)$ code as a partitioned
matrix $\mathbf{H}_{n - k\times n} = [\mathbf{P}_{n-k \times
k}|\mathbf{I}_{n-k \times n-k}]$. $\mathbf{P}$ is a conventional notation for
the sub-matrix that corresponds to information bit positions and $\mathbf{I}$
is an identity matrix that corresponds to parity-check bit positions. Note
that the example ECC code of Equation~\ref{eqn:hg} is in standard form. With
this representation, \mpf{all} codewords take the form $\mathbf{c}_{1\times n}
= [d_0 d_1 ... d_{k-1}|p_0 p_1 ... p_{n-k-1}]$, where $d$ and $p$ are data and
parity-check symbols, respectively.

\subsubsection{Identifying Syndromes Using Miscorrections}
\label{subsubsec:using_miscorrections}

Given that \mpf{on-die ECC conceals error syndromes}, we develop a new
approach for determining the on-die ECC function that \emph{indirectly}
determines error syndromes based on how the ECC mechanism responds when faced
with uncorrectable errors. To induce uncorrectable errors, we deliberately
pause normal DRAM refresh operations \revmp{long enough (e.g., several minutes
at 80$^\circ$C) to cause a large number of data-retention errors (e.g., BER
$>10^{-4}$) throughout a chip. These errors expose a significant number of
miscorrections in different ECC words, and the sheer number of data-retention
errors dominates any unwanted interference from other possible error
mechanisms (e.g., particle strikes~\cite{may1979alpha}).}

\revmp{To control which data-retention errors occur}, we write carefully-crafted
test patterns that restrict \revmp{the} errors to specific bit locations. This
is possible because only cells programmed to the \texttt{CHARGED} state can
experience data-retention errors as discussed in
Section~\ref{subsec:dram_errors}. By restricting pre-correction errors to
certain cells, if a post-correction error is observed in an unexpected location,
it \emph{must} be an artifact of error correction, i.e., a \emph{miscorrection}.
Such a miscorrection is significant since it: (1) signals an error-correction
event, (2) is \emph{purely} a function of the ECC decoding logic, and (3)
indirectly reveals the error syndrome generated by the pre-correction error
pattern. \cro{The indirection occurs because, although the miscorrection does
not expose the raw error syndrome, it \emph{does} reveal that whichever error
syndrome is generated internally by the ECC logic exactly matches the
parity-check matrix column that corresponds to the position of the miscorrected
bit.}

These three properties mean that miscorrections \mpiii{are a reliable tool for
analyzing} ECC functions: for a given pre-correction error pattern, different
ECC functions will generate different error syndromes, and therefore
miscorrections, depending on how the functions' parity-check matrices are
organized. This means that a given ECC function causes miscorrections
\emph{only} within certain bits, and the locations of miscorrection-susceptible
bits differ between functions. Therefore, we can differentiate ECC functions by
\mpiii{identifying which miscorrections are possible for different test patterns.}

\subsubsection{Identifying Useful Test Patterns}
\label{subsubsec:test_patterns}

To construct a set of test patterns that suffice to uniquely identify an ECC
function, we observe that \mpiii{a miscorrection is possible in a
\texttt{DISCHARGED} data bit only if the bit's error syndrome can be produced by
some linear combination of the parity-check matrix columns that correspond to
\texttt{CHARGED} bit locations}. For example, consider the 1-\texttt{CHARGED}
patterns that each set one data bit to the \texttt{CHARGED} state and all others
to the \texttt{DISCHARGED} state. In these patterns, data-retention errors may
\emph{only} occur in either (1) the \texttt{CHARGED} bit or (2) any parity-check
bits that the ECC function also sets to the \texttt{CHARGED} state. With these
restrictions, observable miscorrections may only occur within data bits whose
error syndromes can be created by some linear combination of the parity-check
matrix columns that correspond to the \texttt{CHARGED} cells within the
codeword. 

\cro{As a concrete example, consider the codeword of
Equation~\ref{eqn:codeword}. \texttt{\textbf{C}} and \texttt{\textbf{D}} represent that the
corresponding cell is programmed to the \texttt{CHARGED} and \texttt{DISCHARGED}
states, respectively.}
\vspace{-0.15\baselineskip}
\begin{equation}
	\vspace{-0.25\baselineskip}
	\arraycolsep=1.1pt
	\def\arraystretch{0.8}
	\footnotesize
	\mathbf{c} = \left[
	\begin{array}{cccccccc}	
		\texttt{\textbf{D}}
		& \texttt{\textbf{D}}
		& \textcolor{red}{\texttt{\textbf{C}}}
		& \texttt{\textbf{D}}
		& |
		& \texttt{\textbf{D}}
		& \textcolor{red}{\texttt{\textbf{C}}}
		& \textcolor{red}{\texttt{\textbf{C}}}
	\end{array}
	\right]
\label{eqn:codeword}
\end{equation}
\cro{Because only \texttt{CHARGED} cells \mpi{can} experience data-retention
errors, there are $2^3=8$ possible error syndromes that correspond to the
unique combinations of \texttt{CHARGED} cells failing.
Table~\ref{tab:lincomb} illustrates these eight possibilities. }

\begin{table}[H]
\centering
\arraycolsep=5pt
\def\arraystretch{1}
\footnotesize
$
\begin{array}{cll}
	\textbf{Pre-Correction} & \multirow{2}{*}{~~~~\textbf{Error~Syndrome}} & \textbf{Post-Correction} \\
	\textbf{Error Pattern} &  &  \multicolumn{1}{c}{\textbf{Outcome}} \\
	\hline
	\left[ \arraycolsep=1.5pt \begin{array}{cccccccc} 0 & 0 & 0 & 0 & | & 0 & 0 & 0 \end{array} \right] & \quad\mathbf{0} & \text{No error} \\
	\left[ \arraycolsep=1.5pt \begin{array}{cccccccc} 0 & 0 & 0 & 0 & | & 0 & 0 & \textcolor{red}{1} \end{array} \right] & \quad{\mathbf{H}_{\ast,6}} & \text{Correctable} \\
	\left[ \arraycolsep=1.5pt \begin{array}{cccccccc} 0 & 0 & 0 & 0 & | & 0 & \textcolor{red}{1} & 0 \end{array} \right] & \quad{\mathbf{H}_{\ast,5}} & \text{Correctable} \\
	\left[ \arraycolsep=1.5pt \begin{array}{cccccccc} 0 & 0 & 0 & 0 & | & 0 & \textcolor{red}{1} & \textcolor{red}{1} \end{array} \right] & \quad{\mathbf{H}_{\ast,5} +\mathbf{H}_{\ast,6}} & \textcolor{red}{\text{Uncorrectable}} \\
	\left[ \arraycolsep=1.5pt \begin{array}{cccccccc} 0 & 0 & \textcolor{red}{1} & 0 & | & 0 & 0 & 0 \end{array} \right] & \quad{\mathbf{H}_{\ast,2}} & \text{Correctable} \\
	\left[ \arraycolsep=1.5pt \begin{array}{cccccccc} 0 & 0 & \textcolor{red}{1} & 0 & | & 0 & 0 & \textcolor{red}{1} \end{array} \right] & \quad{\mathbf{H}_{\ast,2} +\mathbf{H}_{\ast,5}} & \textcolor{red}{\text{Uncorrectable}} \\
	\left[ \arraycolsep=1.5pt \begin{array}{cccccccc} 0 & 0 & \textcolor{red}{1} & 0 & | & 0 & \textcolor{red}{1} & 0 \end{array} \right] & \quad{\mathbf{H}_{\ast,2} +\mathbf{H}_{\ast,6}} & \textcolor{red}{\text{Uncorrectable}} \\
	\left[ \arraycolsep=1.5pt \begin{array}{cccccccc} 0 & 0 & \textcolor{red}{1} & 0 & | & 0 & \textcolor{red}{1} & \textcolor{red}{1} \end{array} \right] & \quad{\mathbf{H}_{\ast,2} +\mathbf{H}_{\ast,5} +\mathbf{H}_{\ast,6}} & \textcolor{red}{\text{Uncorrectable}}
\end{array}
$
\caption{\cro{Possible data-retention error patterns, their syndromes, and their
outcomes for the codeword of Equation~\ref{eqn:codeword}.}}
\label{tab:lincomb}
\end{table}

\mpiii{A miscorrection occurs whenever \mpvi{the error syndrome of an
uncorrectable error pattern} matches the parity-check matrix column of a
\emph{non-erroneous} data bit}. In this case, the column's location would then
correspond to the bit position of the miscorrection. However, \mpiii{a
miscorrection only reveals information if it occurs within one of the
\texttt{DISCHARGED} data bits, for only then are we certain that the observed
bit flip is unambiguously a miscorrection rather than an uncorrected
data-retention error}. Therefore, the test patterns we use should maximize the
number of \texttt{DISCHARGED} bits so as to increase the \mpi{number} of
miscorrections \mpvi{that yield information about the ECC function}.

To determine which test patterns to use, we expand upon the \mpvi{approach of
injecting 1-hot errors} described in
Section~\ref{subsubsec:systematically_identifying}. Although we would need to
write data to all codeword bits in order to test every 1-hot error pattern,
on-die ECC does not allow writing directly to the parity-check bits. This leads
to two challenges. First, we cannot test 1-hot error patterns for which the
1-hot error is within the parity-check bits, which means that we cannot
differentiate ECC functions that differ only within their parity-check bit
positions. Fortunately, this is not a problem because, as
Section~\ref{subsubsec:formalizing_ecc_func} discusses in detail, all such
functions are equivalent codes with identical externally-visible
error-correction properties. Therefore, we are free to assume that the
parity-check matrix is in standard form, which \mpvi{specifies parity-check
bits' error syndromes (i.e., $\mathbf{I}_{n-k \times n-k}$) and obviates the
need to experimentally determine them.} 

Second, \mpi{writing the $k$ bits of the dataword with a single \texttt{CHARGED}
cell results in a codeword with an \emph{unknown} number of \texttt{CHARGED}
cells because the ECC function independently determines the values of remaining
$n-k$ parity-check bits. As a result, the final codeword may contain anywhere
from $1$ to $n-k+1$ \texttt{CHARGED} cells, and the number of \texttt{CHARGED}
cells will vary for different test patterns. Because we cannot directly access
the parity-check bits' values, we do not know which cells are \texttt{CHARGED}
for a given test pattern, and therefore, we cannot tie post-correction errors
back to particular pre-correction error patterns. Fortunately, we can work
around this problem by considering \emph{all possible} error patterns that a
given codeword can experience, which amounts to examining all combinations of
errors that the \texttt{CHARGED} cells can experience. Table~\ref{tab:lincomb}
illustrates this for when the dataword is programmed with a 1-\texttt{CHARGED}
test pattern (as shown in Equation~\ref{eqn:codeword}). In this example, the
encoded codeword contains three \texttt{CHARGED} cells, which may experience any
of $2^3$ possible error patterns.
Section~\ref{subsubsec:testing_charged_patterns} discusses how we can accomplish
testing all possible error patterns in practice by exploiting the fact that
data-retention errors occur uniform-randomly, so testing across many different
codewords provides samples from many different error patterns \mpvi{at once}.}

\subsubsection{\mpi{Shortened Codes}}

\mpi{Linear block codes can be either of} \emph{full-length} if all possible
error syndromes are present within the parity-check matrix (e.g., all $2^p - 1$
error syndromes for a Hamming code with $p$ parity-check bits\mpi{, as is the
case for the code shown in Equation~\ref{eqn:hg}}) or \emph{shortened} if one or
more information symbols are truncated while retaining the same number of
parity-check symbols~\cite{costello1982error, huffman2010fundamentals}.
\mpi{This distinction is crucial for determining appropriate test patterns
because, for full-length codes,} the 1-\texttt{CHARGED} patterns identify the
miscorrection-susceptible bits for all possible error syndromes. In this case,
testing additional patterns that have more than one \texttt{CHARGED} bit
provides no new information because any resulting \mpi{error syndromes are
already tested} using the 1-\texttt{CHARGED} patterns.

However, for \emph{shortened codes}, the 1-\texttt{CHARGED} patterns may not
provide enough information to uniquely identify the ECC function because the
1-\texttt{CHARGED} patterns can no longer test for the missing error syndromes.
Fortunately, we can recover the \mpt{missing information by reconstructing the
truncated error syndromes using pairwise \emph{combinations} of the
1-\texttt{CHARGED} patterns. For example, asserting two \texttt{CHARGED} bits
effectively tests an error syndrome that is the linear combination of the bits'
corresponding parity-check matrix columns. Therefore, by supplementing the
1-\texttt{CHARGED} patterns with the 2-\texttt{CHARGED} patterns, we effectively
encompass the error syndromes that were shortened.}
Section~\ref{subsec:beer1eval_correctness} evaluates BEER's sensitivity to code
length, showing that the 1-\texttt{CHARGED} patterns are indeed sufficient for
full-length codes and the \{1,2\}-\texttt{CHARGED} patterns for shortened codes
\mpi{that we evaluate \mpii{with} dataword lengths \mpii{between} 4 and 247}.

%% file: 5_beer.tex
\section{Bit-Exact Error Recovery (BEER)}
\label{sec:beeri}

Our goal in this work is to develop a methodology that reliably and accurately
determines the full ECC function (i.e., its parity-check matrix) for any DRAM
on-die ECC implementation without requiring hardware \mpi{tools, prerequisite
knowledge about the DRAM chip or on-die ECC mechanism,} or access to \mpf{ECC
metadata (e.g., error syndromes, parity information)}. To this end, we present
BEER, which systematically determines the ECC function by observing how it
reacts when subjected to carefully-crafted uncorrectable error patterns. BEER
implements the ideas developed throughout Section~\ref{sec:determining_ecc_func}
and consists of three key steps: (1) experimentally inducing miscorrections, (2)
analyzing observed post-correction errors, and (3) solving for the ECC function.

\mpt{This section describes each of these steps in detail in the context of
experiments using 32, 20, and 28 real LPDDR4 DRAM chips from three major
manufacturers, whom we anonymize for confidentiality reasons as A, B, and C,
respectively. We perform all tests using a temperature-controlled
infrastructure with precise control over the timings of refresh and other DRAM
bus commands.}

\subsection{Step 1: Inducing Miscorrections}

\cro{To induce miscorrections as discussed in
Section~\ref{subsubsec:test_patterns}, we must first identify the (1)
\texttt{CHARGED} and \texttt{DISCHARGED} encodings of each cell and (2) layout
of individual datawords within the address space. This section describes how we
determine these in a way that is applicable to any DRAM chip.}

\subsubsection{Determining \texttt{CHARGED} and \texttt{DISCHARGED} States} 
\label{subsubsec:tacd_layout}

We determine the encodings of the \texttt{CHARGED} and \texttt{DISCHARGED}
states by experimentally measuring the layout of true- and anti-cells throughout
the address space as done in prior works~\cite{kim2014flipping,
kraft2018improving, patel2019understanding}. \cro{We write data `0' and data `1'
test patterns to the entire chip while pausing DRAM refresh for 30 minutes at
temperatures between $30-80^\circ$C.} The resulting data-retention error
patterns reveal the true- and anti-cell layout since each test pattern isolates
one of the cell types. We find that chips from manufacturers A and B use
exclusively true-cells, and chips from manufacturer C use 50\%/50\%
true-/anti-cells organized in alternating \cro{blocks of rows with block lengths
of 800, 824, and 1224 rows. These observations are consistent with the results
of similar experiments performed by prior work~\cite{patel2019understanding}.}

\subsubsection{Determining the Layout of Datawords} 
\label{subsubsec:ecc_word_layout}

To determine which \mpt{addresses} correspond to individual ECC datawords, we
program \mpt{one cell per row\footnote{\mpt{We assume that ECC words do not
straddle row boundaries since accesses would then require reading two rows
simultaneously. However, one cell per \emph{bank} can be tested to accommodate
this case if required.}} to the \texttt{CHARGED} state with all other cells
\texttt{DISCHARGED}.} We then sweep the refresh window $t_{REFw}$ from 10
seconds to 10 minutes at $80^\circ$C \mpi{to induce uncorrectable errors.
Because \emph{only} \texttt{CHARGED} cells can fail, post-correction errors may
\emph{only} occur in bit positions corresponding to either} (1) the
\texttt{CHARGED} cell itself or (2) \mpi{\texttt{DISCHARGED} cells due to a
miscorrection.} By \mpi{sweeping the bit position of the \texttt{CHARGED} cell
within the dataword}, we observe miscorrections that are restricted exclusively
to \emph{within the same ECC dataword.} We find that chips from all three
manufacturers use identical ECC word layouts: each contiguous 32B region of DRAM
comprises two 16B ECC words that are interleaved at byte granularity. A 128-bit
dataword is consistent with prior industry and academic works on on-die
ECC~\cite{kwak2017a, kwon2017an, micron2017whitepaper, patel2019understanding}.

\subsubsection{Testing With {1,2}-\texttt{CHARGED} Patterns} 
\label{subsubsec:testing_charged_patterns}

To test each of the 1- or 2-\texttt{CHARGED} patterns, we program an equal
number of datawords with each test pattern. For example, a 128-bit dataword
yields ${128\choose1} = 128$ and ${128\choose2} = 8128$ 1- and
2-\texttt{CHARGED} test patterns, respectively. \mpi{As
Section~\ref{subsubsec:test_patterns} discusses, BEER must identify all possible
miscorrections for each test pattern. To do so, BEER must exercise all possible
error patterns that a codeword programmed with a given test pattern can
experience (e.g., up to $2^{10} = 1024$ unique error patterns for a (136, 128)
Hamming code using a 2-\texttt{CHARGED} pattern).}

\mpi{Fortunately, although BEER must test a large number of error patterns, even
a single DRAM chip typically contains millions of ECC words (e.g., $2^{24}$
128-bit words for a 16 Gib chip), and we simultaneously test them all when we
reduce the refresh window across the entire chip. Because data-retention errors
occur uniform-randomly (discussed in Section~\ref{subsec:dram_errors}), every
ECC word tested provides an independent sample of errors. Therefore, even one
experiment provides millions of samples of different error patterns within the
\texttt{CHARGED} cells, and running multiple experiments at different operating
conditions (e.g., changing temperature or the refresh window) across multiple
DRAM chips\footnote{\mpi{Assuming chips of the same model use the same on-die
ECC mechanism, which our experimental results in
Section~\ref{subsubsec:testing_charged_patterns} support.}} dramatically
increases the sample size, making the probability of not observing a given error
pattern exceedingly low.. We analyze experimental runtime in
Section~\ref{subsec:experiment_runtime}.}

Table~\ref{tab:misc_profiles} \mpii{illustrates testing the 1-\texttt{CHARGED}
patterns using the ECC function given by Equation~\ref{eqn:hg}. There are four
test patterns, and Table~\ref{tab:misc_profiles} shows the miscorrections that
are possible for each one} assuming that all cells are true cells. For this ECC
function, miscorrections are possible \mpiii{\emph{only}} for test pattern 0,
\mpiii{and} no pre-correction error pattern exists that can cause miscorrections
for the other test patterns. Note that, for errors in the \texttt{CHARGED}-bit
positions, we cannot be certain whether a post-correction error is a
miscorrection or simply a data-retention error, so we label it using `?'. We
refer to the cumulative pattern-miscorrection pairs as a \emph{miscorrection
profile}\mpiii{. Thus, Table~\ref{tab:misc_profiles} shows the miscorrection
profile of} the ECC function given by Equation~\ref{eqn:hg}.

\begin{table}[h]
    \scriptsize
    \setlength\tabcolsep{3pt}
    \def\arraystretch{1}
    \begin{tabular}{c|c|c}
        \textbf{1-\texttt{CHARGED} Pattern ID} & \multirow{2}{*}{\textbf{1-\texttt{CHARGED} Pattern}} & \multirow{2}{*}{\textbf{Possible Miscorrections}} \\ 
        \textbf{(Bit-Index of \texttt{CHARGED} Cell)} & & \\ \hline
        3 & $\left[\texttt{\textbf{D}}~\texttt{\textbf{D}}~\texttt{\textbf{D}}~\textcolor{red}{\texttt{\textbf{C}}}\right]$
        & $\left[{-}~{-}~{-}~?\right]$ \\
        2 & $\left[\texttt{\textbf{D}}~\texttt{\textbf{D}}~\textcolor{red}{\texttt{\textbf{C}}}~\texttt{\textbf{D}}\right]$
        & $\left[{-}~{-}~?~{-}\right]$ \\
        1 & $\left[\texttt{\textbf{D}}~\textcolor{red}{\texttt{\textbf{C}}}~\texttt{\textbf{D}}~\texttt{\textbf{D}}\right]$
        & $\left[{-}~?~{-}~{-}\right]$ \\
        0 & $\left[\textcolor{red}{\texttt{\textbf{C}}}~\texttt{\textbf{D}}~\texttt{\textbf{D}}~\texttt{\textbf{D}}\right]$
        & $\left[?~\textcolor{red}{\texttt{\textbf{1}}}~\textcolor{red}{\texttt{\textbf{1}}}~\textcolor{red}{\texttt{\textbf{1}}}\right]$ \\
    \end{tabular}
    \caption{\mpi{Example miscorrection profile for the ECC function given in Equation~\ref{eqn:hg}.}}
    \label{tab:misc_profiles}
\end{table}

\begin{figure}[h]
    \centering
    \includegraphics[width=\linewidth]{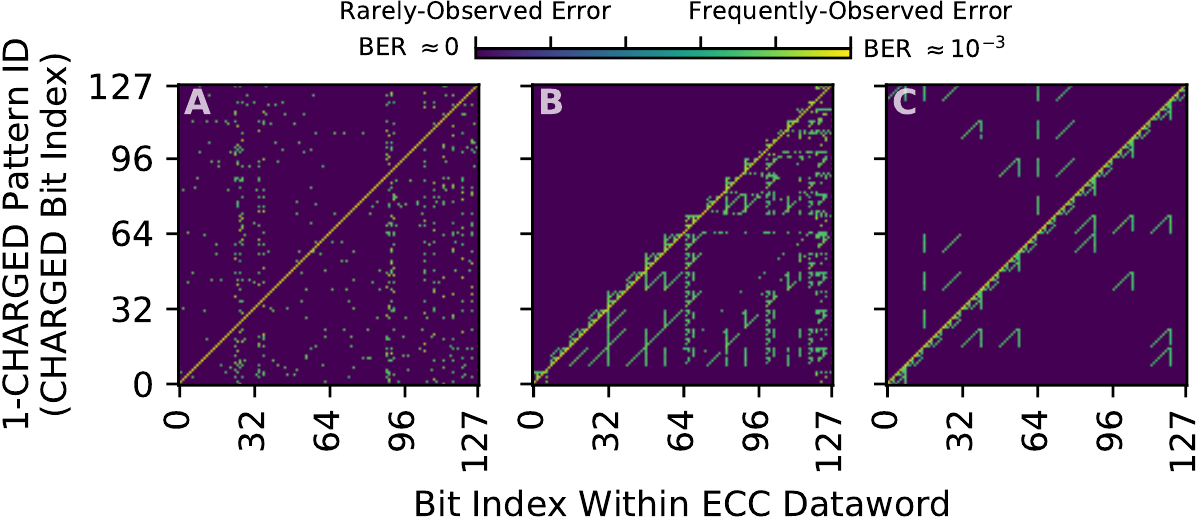}
    \caption{\mpi{Errors observed in a single representative chip from each
    manufacturer using the 1-\texttt{CHARGED} test patterns, showing that
    manufacturers appear to use different ECC functions.}}
    \label{fig:misc_profiles}
\end{figure}

To obtain the miscorrection \mpiii{profile of the on-die ECC function within
each DRAM chip that we test}, we lengthen the refresh window $t_{REFw}$ to
between 2 minutes, where uncorrectable errors begin to occur \mpi{frequently
(BER $\approx 10^{-7}$),} and 22 minutes, where nearly all ECC words exhibit
uncorrectable errors \mpi{(BER $\approx 10^{-3}$)}, in 1 minute intervals at
80$^\circ$C. \mpii{During each experiment, we record which bits are susceptible
to miscorrections for each test pattern} \mpi{(analogous to
Table~\ref{tab:misc_profiles})}. \mpii{Figure~\ref{fig:misc_profiles} shows this
information graphically, giving} the logarithm of the number of errors observed
in each bit position ($X$-axis) for each 1-\texttt{CHARGED} test pattern
($Y$-axis). \mpi{The data is taken from the true-cell regions of a single
representative chip from each manufacturer. \mpii{Errors} in the
\texttt{CHARGED} bit positions (i.e., \mpi{where} $Y=X$) stand out clearly
because \mpiii{they occur alongside all miscorrections as uncorrectable errors.}}

The data shows that miscorrection profiles vary significantly between different
manufacturers. \mpiii{This is likely because each manufacturer uses} a different
parity-check matrix: the possible miscorrections for a given test pattern depend
on which parity-check matrix columns are used to construct error syndromes. With
different matrices, different columns combine to form different error syndromes.
The miscorrection profiles of manufacturers B and C exhibit repeating patterns,
which \mpi{likely} occur due to regularities in how syndromes are organized in
the parity-check matrix, whereas the matrix of manufacturer A appears to be
relatively unstructured. We suspect that manufacturers use different ECC
functions because each manufacturer employs their own circuit design, and
specific parity-check matrix organizations lead to more favorable circuit-level
tradeoffs (e.g., layout area, critical path lengths).

We find that chips of the same model number from the same manufacturer yield
identical miscorrection profiles, which (1) validates that we are observing
design-dependent data and (2) confirms that chips from the same manufacturer and
product generation appear to use the same ECC functions. To sanity-check
our results, we use EINSim~\cite{patel2019understanding, eccsimgithub} to
simulate the miscorrection profiles of the final parity-check matrices we obtain
from our experiments with real chips, and we \mpi{observe} that \mpiii{the
miscorrection profiles obtained via simulation match those measured via real
chip experiments.}

\subsection{Step 2: Analyzing Post-Correction Errors}
\label{subsec:analyzing_postecc_errors}

In practice, BEER may either (1) fail to observe a possible miscorrection or
(2) misidentify a miscorrection due to \mpi{unpredictable} transient errors
(e.g., soft errors from particle strikes, variable-retention time errors,
voltage fluctuations). These events can theoretically pollute the miscorrection
profile with \mpi{incorrect data}, potentially resulting in an \emph{illegal}
miscorrection profile, \mpi{i.e., one} that does not match \emph{any} ECC
function.

Fortunately, case (1) is unlikely given the sheer number of ECC words even a
single chip provides for testing (discussed in
Section~\ref{subsubsec:testing_charged_patterns}). While it is possible that
different ECC words throughout a chip use different ECC functions, we believe
that this is unlikely because it \mpi{complicates the design with no clear
benefits}. Even if a chip does use more than one ECC function, the different
functions will likely follow patterns aligning with DRAM substructures (e.g.,
alternating between DRAM rows or subarrays~\cite{kim2012case, kim2018solar}),
and we can test each region individually. 

\begin{figure}[h]
    \centering
    \includegraphics[width=\linewidth]{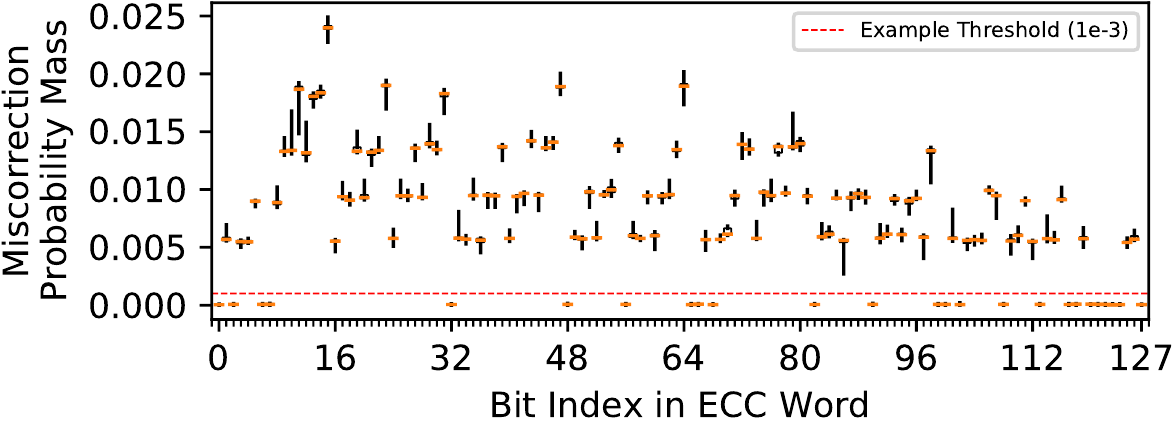}
    \caption{\mpi{Relative probability of observing a miscorrection in each bit
position aggregated across all 1-\texttt{CHARGED} test patterns for a
representative chip of manufacturer B. \mpi{The dashed line shows a threshold
filter separating zero and nonzero values.}}}
    \label{fig:misc_probs_B}
\end{figure}

\mpi{Similarly, case (2) is unlikely because transient errors occur randomly and
rarely~\cite{qureshi2015avatar} as compared with the data-retention error rates
that we induce for BEER \mpi{($>10^{-7}$)}, so transient error
occurrence counts are far lower than those of real miscorrections that are
observed frequently in miscorrection-susceptible bit positions. Therefore, we
apply a simple threshold filter to remove rarely-observed post-correction errors
from the miscorrection profile.} Figure~\ref{fig:misc_probs_B} shows the
\mpi{relative} probability of observing a miscorrection in each bit position
aggregated across all 1-\texttt{CHARGED} test patterns for a representative
chip from manufacturer B. Each data point is a boxplot that shows the full
distribution \mpi{of probability values, i.e., min, median, max, and
interquartile-range (IQR), observed when sweeping the refresh window from 2 to
22 minutes (i.e., the same experiments described in
Section~\ref{subsubsec:testing_charged_patterns}).}

We see that zero and nonzero probabilities are distinctly separated, \mpt{so we
can} robustly resolve miscorrections for each bit. Furthermore, each
distribution is extremely tight, meaning that any of the individual experiments
(i.e., any single component of the distributions) is suitable for identifying
miscorrections. Therefore, a simple threshold filter \mpi{(illustrated in
Figure~\ref{fig:misc_probs_B})} distinctly separates \mpvi{post-correction
errors} that occur near-zero times \mpvi{from miscorrections} that occur
significantly more often.

\subsection{Step 3: Solving for the ECC Function}
\label{subsubsec:rev_engr_ecc_function}

We use the Z3 SAT solver~\cite{de2008z3} (described in
Section~\ref{subsec:sat_solver}) to identify the exact ECC function given a
miscorrection profile. To determine the encoding ($F_{encode}$) and decoding
($F_{decode}$) functions, we express them as unknown generator ($\textbf{G}$)
and parity-check ($\textbf{H}$) matrices, respectively. We then add the
following constraints to the SAT solver for $\textbf{G}$ and $\textbf{H}$:
\begin{enumerate}
\item Basic linear code properties (e.g., unique $\textbf{H}$ columns).
\item Standard form matrices, as described in Section~\ref{subsubsec:formalizing_ecc_func}.
\item Information contained within the miscorrection profile
(i.e., pattern $i$ can(not) yield a miscorrection in bit $j$).
\end{enumerate}

\noindent
Upon evaluating the SAT solver with these constraints, the resulting
$\textbf{G}$ and $\textbf{H}$ matrices represent the ECC encoding and decoding
functions, respectively, that cause the observed miscorrection profile. To
verify that no other ECC function may result in the same miscorrection
profile, we simply repeat the SAT solver evaluation with the additional
constraint that the already discovered $\textbf{G}$ and $\textbf{H}$ matrices
are invalid. If the SAT solver finds another ECC function that satisfies the
new constraints, the solution is not unique. 

\cro{To \mpii{seamlessly} apply BEER to the DRAM chips that we test, we develop an
open-source C++ application~\cite{beergithub} that incorporates the SAT solver
and determines the ECC function corresponding to an arbitrary miscorrection
profile. The tool exhaustively searches for all possible ECC functions that
satisfy the aforementioned constraints and therefore will generate the input
miscorrection profile.} Using this tool, we apply BEER to miscorrection profiles
that we experimentally measure across all chips using refresh windows up to 30
minutes and temperatures up to 80$^\circ$C. We find that BEER uniquely
identifies the ECC function for all manufacturers. Unfortunately, we are unable
to publish the resulting ECC functions for confidentiality reasons as set out in
Section~\ref{subsec:motivation_secrecy}. \cro{Although we are confident in our
results because our SAT solver tool identifies a unique ECC function that
explains the observed miscorrection profiles for each chip, we have no way to
validate BEER's results against a groundtruth. To overcome this limitation, we
demonstrate BEER's correctness using simulation in
Section~\ref{subsec:beer1eval_correctness}.}

\subsection{Requirements and Limitations}
\label{subsec:beeri_limitations}

\mpii{Although we demonstrate BEER's effectiveness using both experiment and
simulation, BEER has several testing requirements and limitations that we review
in this section.}

\noindent
\textbf{\mpii{Testing Requirements}}

\begin{itemize}
\item \revmpf{\emph{Single-level ECC}: BEER assumes \mpvi{that there is no
second level of ECC (e.g., rank-level ECC in the DRAM controller) present}
during testing.\footnote{\revmpf{We can potentially extend BEER to multiple
levels of ECC by extending the SAT problem to the concatenated code formed by
the combined ECCs and constructing test patterns that target each level
sequentially, but \mpi{we leave this direction to future work}.}} This is
reasonable since system-level ECCs can typically be bypassed (e.g., \mpi{via}
FPGA-based testing \mpi{or disabling through the BIOS) or
reverse-engineered~\cite{cojocar19exploiting}, even in the presence of on-die
ECC, before applying BEER}.}

\item \mpii{\emph{Inducing data-retention errors:}} BEER requires finding a
refresh window (i.e., $t_{REFw}$) for each chip that is long enough to
\mpii{induce data-retention errors and} expose miscorrections. Fortunately, we
find that \mpii{refresh windows} between 1-30 minutes at 80$^\circ$C
\mpii{reveal} more than enough miscorrections to apply BEER. In general, the
refresh window \mpi{can be} easily modified (discussed in
Section~\ref{subsec:dram_errors}), \mpii{and because data-retention errors are
fundamental to DRAM technology,} BEER applies to all DDRx DRAM families
regardless of their data access protocols and will likely hold for future DRAM
chips\mpiii{, whose data-retention error rates will likely be even more
prominent~\cite{micron2017whitepaper, kang2014co, nair2016xed, gong2017dram,
kwon2014understanding, meza2015revisiting, oh20153, kim2007low, son2015cidra,
liu2013experimental}.}

\end{itemize}

\noindent
\textbf{\revmp{Limitations}}

\begin{itemize}
\item \emph{ECC code type}: BEER works on systematic linear block codes, which
are \mpi{commonly employed} for latency-sensitive main memory chips since: (i) they
allow the data to be directly accessed without additional
operations~\cite{zhang2015vlsi} and (ii) stronger codes (e.g.,
LDPC~\cite{gallager1963low}, concatenated codes~\cite{forney1965concatenated})
cost considerably more area and latency~\cite{cai2017error, nishi2019advances}.

\item \emph{No groundtruth}: BEER alone cannot confirm whether the ECC function
that it identifies is the correct answer. However, if BEER finds exactly one ECC
function that explains the experimentally observed miscorrection profile, it is
\mpi{very} likely that the ECC function is correct.

\item \emph{\mpi{Disambiguating equivalent codes}}: On-die ECC does not expose
the parity-check bits, so BEER can only determine the ECC function to an
equivalent code \mpi{(discussed in Sections~\ref{subsubsec:formalizing_ecc_func}
and~\ref{subsubsec:test_patterns}). Fortunately, equivalent codes differ only in
their internal metadata representations, so this limitation should not hinder
most third-party studies. In general,} we are unaware of any way to disambiguate
equivalent codes without accessing the ECC mechanism's internals.

\end{itemize}

%% file: 6_beer_eval.tex
\section{BEER Evaluation}
\label{sec:beer1eval}

\cro{We evaluate BEER's correctness in simulation, \mpi{SAT solver} performance
on a real system, and experimental runtime analytically. Our evaluations both
(1) show that BEER is practical and correctly identifies the ECC function within
our simulation-based analyses, and (2) provide intuition for how the SAT
problem's complexity scales for longer ECC codewords.}

\subsection{Simulation-Based Correctness Evaluation}
\label{subsec:beer1eval_correctness}

We simulate applying BEER to DRAM chips with on-die ECC using a modified version
of the EINSim~\cite{patel2019understanding, eccsimgithub} open-source DRAM
error-correction simulator that we also publicly release~\cite{eccsimgithub}. We
simulate \mpi{115300} single-error correction Hamming code functions that are
representative of those used for on-die ECC~\cite{im2016im, nair2016xed,
micron2017whitepaper, micron2019whitepaper, oh20153, kwak2017a, kwon2017an,
patel2019understanding}: 2000 each for dataword \mpiii{lengths} between 4 and 57 bits,
\mpii{100 each between 58 and 120 bits, and 100 each for selected values}
between 121 and 247 bits because longer codes require significantly more
simulation time. For each ECC function, we simulate inducing data-retention
errors within the 1-, 2-, and 3-\texttt{CHARGED}\footnote{We include the
3-\texttt{CHARGED} patterns to show that \mpvii{they fail to uniquely identify
all ECC functions despite comprising combinatorially more test patterns than the
combined 1- and 2-\texttt{CHARGED} patterns.}} test patterns according to the
data-retention error properties outlined in Section~\ref{subsec:dram_errors}.
For each test pattern, we \mpiv{model a real experiment by simulating} $10^9$
ECC words and \mpiii{data-retention error rates} ranging from $10^{-5}$ to
$10^{-2}$ to obtain a miscorrection profile. \cro{Then, we apply BEER to the
miscorrection profiles and show that BEER correctly recovers the original ECC
functions.}

\begin{figure}[h]
    \centering
    \includegraphics[width=\linewidth]{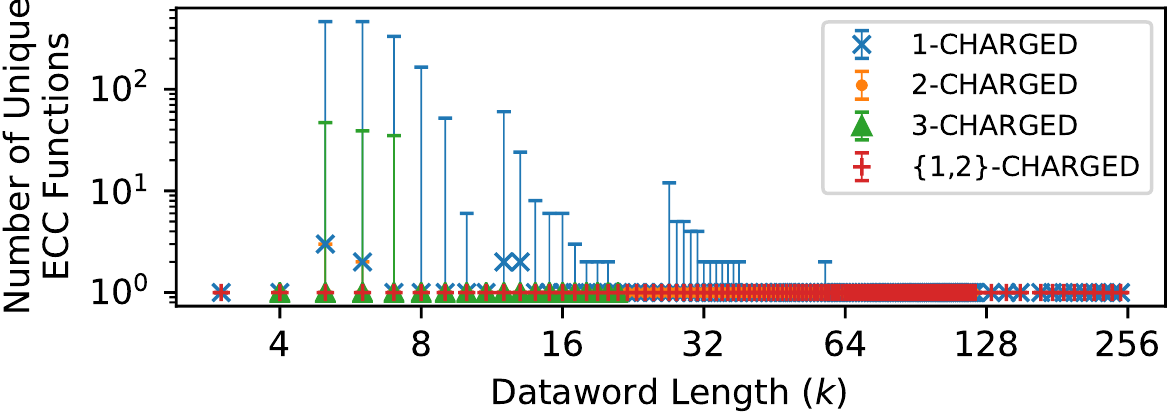}
    \caption{Number of ECC functions that match miscorrection profiles created
    using different test patterns.}
    \label{fig:beer_correctness}
\end{figure}

\mpo{Figure~\ref{fig:beer_correctness} shows how many unique ECC functions BEER
finds when using different test patterns to generate miscorrection profiles. For
each dataword length tested, we show the minimum, median, and maximum number of
solutions identified across all miscorrection profiles. The data shows that BEER
is always able to recover the original \mpi{unique} ECC function using the
\{1,2\}-\texttt{CHARGED} configuration that \mpt{uses both the
1-\texttt{CHARGED} and 2-\texttt{CHARGED} test patterns}. \mpf{For full-length
codes (i.e., \mpi{with dataword lengths} $k\in{4, 11, 26, 57, 120, 247,...}$)
that contain all possible error syndromes within the parity-check matrix by
construction, all test patterns uniquely determine the ECC function, including
the 1-\texttt{CHARGED} patterns alone.}}

On the other hand, the individual 1-, 2-, and 3-\texttt{CHARGED} patterns
sometimes identify multiple ECC functions for shortened codes, with more
solutions identified both for (1) shorter codes and (2) codes with
more aggressive shortening. However, the data shows that BEER often still
uniquely identifies the ECC function even using \mpi{only} the
1-\texttt{CHARGED} patterns \mpii{(i.e., for 87.7\% of all codes simulated)} and
\emph{always} does so with the \{1,2\}-\texttt{CHARGED} patterns. This is
consistent with the fact that shortened codes expose fewer error syndromes to
test (discussed in Section~\ref{subsubsec:test_patterns}). It is important to
note that, \mpv{even if BEER identifies} multiple solutions, it still
narrows a combinatorial-sized search space to a tractable number of ECC
functions that are \mpii{well suited} to more expensive analyses (e.g.,
intrusive error-injection, die imaging techniques, or manual inspection).

While our simulations do not model interference from transient errors, such
errors are rare events~\cite{qureshi2015avatar} when compared with the amount of
\mpiv{uncorrectable data-retention errors that BEER induces}. Even if sporadic
transient errors were to occur, Section~\ref{subsec:analyzing_postecc_errors}
discusses in detail how BEER mitigates their impact on the miscorrection profile
using a simple thresholding filter.

\subsection{Real-System Performance Evaluation}
\label{subsec:beer1eval_perf}

\begin{figure}[h]
    \centering
    \includegraphics[width=\linewidth]{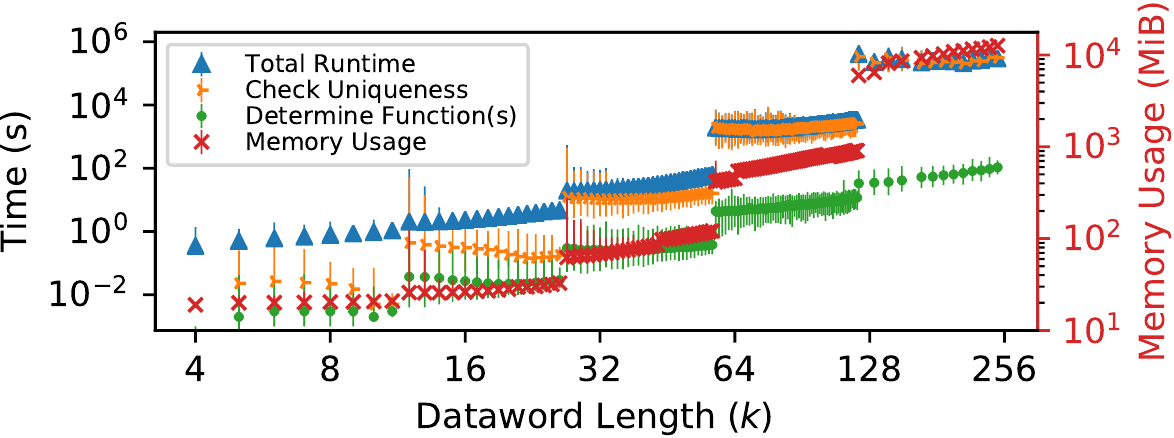}
    \caption{Measured BEER runtime (left y-axis) and memory usage (right
    y-axis) for different ECC codeword lengths.}
    \label{fig:beer1_perf}
\end{figure}

We evaluate BEER's performance and memory usage using ten servers with 24-core
2.30 GHz Intel Xeon(R) Gold 5118 CPUs~\cite{intel2020xeongold5118} and 192 GiB
2666 MHz DDR4 DRAM~\cite{jedec2012ddr4} \mpt{each}. All measurements are taken
\mpi{with Hyper-Threading~\cite{intel2020xeongold5118} enabled and all cores
fully occupied}. Figure~\ref{fig:beer1_perf} shows \mpii{overall runtime} and
memory usage when running BEER with the 1-\texttt{CHARGED} patterns for
different ECC code lengths on a log-log plot \mpii{along with} the time required
to (1) solve for the ECC function (``Determine Function'') and (2) verify the
uniqueness of the solution (``Check Uniqueness''). Each data point gives the
minimum, median, and maximum values observed across our simulated ECC
functions (described in Section~\ref{subsec:beer1eval_correctness}). We see
that the total runtime and memory usage are negligible for short codes and grow
as large as 62 hours and 11.4 GiB of memory for large codes. \mpii{For a
representative dataword length of 128 bits, the median total runtime and memory
usage are 57.1 hours and 6.3 GiB, respectively.} At each code length where we
add an additional parity-check bit, the runtime and memory usage jump
accordingly since the complexity of the SAT evaluation problem increases by an
extra dimension.

The total runtime is quickly dominated by the SAT solver checking uniqueness,
which requires exhaustively exploring the entire search space of a given ECC
function. However, simply determining the solution ECC function(s) is much
faster, requiring less than 2.7 minutes even for the longest codes evaluated and
for shortened codes that potentially have multiple solutions \mpi{using only}
the 1-\texttt{CHARGED} patterns. From this data, we conclude that BEER is
practical for reasonable-length codes used for on-die ECC (e.g., $k=64, 128$).
However, our BEER implementation has room for optimization, e.g., using
dedicated GF(2) BLAS libraries (e.g., LELA~\cite{hovinen2011lela}) or advanced
SAT solver theories (e.g., SMT bitvectors~\cite{brummayer2009boolector}), and an
optimized implementation would likely improve performance, enabling BEER's
application to an even greater range of on-die ECC functions.
Section~\ref{subsec:future_work} discusses such optimizations in greater detail.
Nevertheless, BEER is a one-time \emph{offline} process, so it need not be
aggressively performant in most use-cases.

\subsection{Analytical Experiment Runtime Analysis}
\label{subsec:experiment_runtime}

Our experimental runtime \mpi{is overwhelmingly bound by waiting for
data-retention errors to occur during a lengthened refresh window (e.g., 10
minutes) while} interfacing with the DRAM chip requires \mpi{only on the order
of milliseconds} (e.g., 168 ms to read an entire 2 GiB LPDDR4-3200
chip~\cite{jedec2014lpddr4}). Therefore, we estimate \mpi{total} experimental
runtime \mpi{as the sum of the refresh windows that we individually test}. For
the data we present in Section~\ref{subsubsec:testing_charged_patterns},
\mpi{testing each refresh window between 2 to 22 minutes in 1 minute increments
requires a combined 4.2 hours of testing \mpii{for a single chip}. However, if
chips of the same model number use the same ECC functions (as \mpii{our data
supports in} Section~\ref{subsubsec:testing_charged_patterns}), \mpii{we can
reduce overall testing latency by parallelizing individual tests} across
different chips.} Furthermore, because BEER is \mpi{likely} a one-time exercise
for a given DRAM chip, it is sufficient that BEER is practical \mpii{offline}.

%% file: 7_use_cases.tex
\begin{figure*}[b]
    \vspace*{-2ex}
    \centering
    \includegraphics[width=\linewidth]{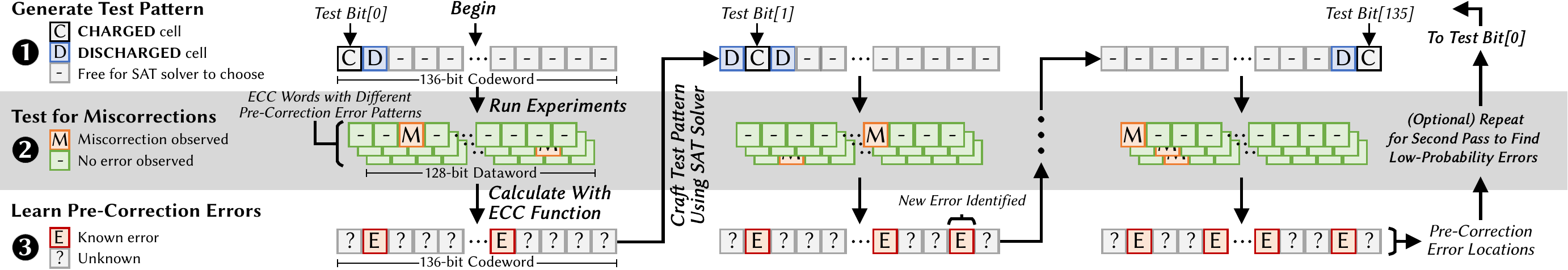}
    \caption{\mpi{Example of running BEEP on a single \mpii{136-bit ECC
    codeword} to identify locations of pre-correction errors.}}
    \label{fig:beep_steps}
\end{figure*}

\section{Example Practical Use-Cases}
\label{section:usecases}

BEER empowers third-party DRAM users to decouple the reliability
characteristics of modern DRAM chips from any particular on-die ECC function
that a chip implements. This section discusses five concrete analyses that
BEER enables. To our knowledge, BEER is the first work capable of
inferring this information without bypassing the on-die ECC mechanism. We hope
that end users and future works find more ways to extend and apply BEER
in practice.

\subsection{BEEP: Profiling for \mpt{Raw Bit} Errors}
\label{subsec:beep}

We introduce \underline{B}it-\underline{E}xact \underline{E}rror
\underline{P}rofiling (BEEP), a new data-retention error profiling algorithm
enabled by BEER that infers the number and bit-exact locations of
\mpt{pre-correction} error-prone cells when given a set of operating conditions that
cause uncorrectable errors in an ECC word. To our knowledge, BEEP is the first
DRAM error profiling methodology capable of identifying bit-exact error
locations throughout the entire on-die ECC codeword, including within the
parity bits.

\subsubsection{BEEP: Inference Based on Miscorrections}

Because miscorrections are purely a function of the ECC logic (discussed in
Section~\ref{subsubsec:using_miscorrections}), an observed miscorrection
indicates that a specific pre-correction error pattern has occurred. Although
several such patterns can map to the same miscorrection, BEEP narrows down the
possible pre-correction error locations by using the known parity-check matrix
\jkz{(after applying BEER)} to construct test patterns for additional
experiments that disambiguate the possibilities. At a high level, BEEP crafts
test patterns to reveal errors as it incrementally traverses each codeword bit,
\mpiv{possibly using multiple passes to capture low-probability errors}. As BEEP
iterates over the codeword, it builds up a list of suspected error-prone cells.

\mpiii{BEEP comprises three} phases: \circled{1} \jkz{crafting suitable test
patterns}, \circled{2} experimental testing \jkz{with crafted patterns}, and
\circled{3} calculating pre-correction error locations from observed
miscorrections. Figure~\ref{fig:beep_steps} illustrates these three phases in an
example where BEEP profiles for pre-correction errors in a 128-bit ECC dataword.
The following sections explain each of the three phases and refer to
Figure~\ref{fig:beep_steps} as a running example.

\subsubsection{Crafting Suitable Test Patterns}
\label{subsubsec:beep_test_patterns}

Conventional DRAM error profilers (e.g.,~\mpiv{\cite{venkatesan2006retention,
liu2012raidr, liu2013experimental, khan2016parbor, lee2017design,
patel2017reaper, khan2017detecting, cheng2002neighborhood, van1991testing,
kraft2018improving, hassan2017softmc, jha2003testing}}) use carefully designed
test patterns that induce worst-case circuit conditions in order to maximize
their coverage of potential errors~\cite{adams2002high, mrozek2019multi}.
\cro{Unfortunately, on-die ECC encodes all data into codewords, so the intended
software-level test patterns likely do not maintain their carefully-designed
properties when written to the physical DRAM cells.} BEEP circumvents these
ECC-imposed restrictions by using a SAT solver along with the known ECC function
\jkz{(via BEER)} to craft test patterns that \mpvi{both} (1) \emph{locally} induce the
worst-case circuit conditions and (2) result in \emph{observable miscorrections}
if suspected error-prone cells do indeed fail.

Without loss of generality, we assume that the worst-case conditions for a given
bit occur when its neighbors are programmed with the opposite charge states,
which prior work shows to exacerbate circuit-level coupling effects and increase
error rates~\mpiv{\cite{adams2002high, liu2013experimental, khan2016parbor,
mrozek2019multi, van2002address, redeker2002investigation, al2004effects,
seyedzadeh2017mitigating, konishi1989analysis, li2011dram, nakagome1988impact}}.
If the design of a worst-case pattern is not known, or if it has a different
structure than we assume, BEEP \mpi{can be adapted} by simply modifying the
relevant SAT solver constraints (described below). To ensure that BEEP observes
a miscorrection when a given error occurs, BEEP crafts a pattern that will
suffer a miscorrection if the error occurs \emph{alongside an
already-discovered} error. We express these conditions to the SAT solver using
the following constraints:

\begin{enumerate}
\item Bits adjacent to the target bit \mpt{have opposing charge states}.
\item One or more miscorrections is possible using some combination of the
already-identified data-retention errors.
\end{enumerate}

\noindent
\cro{Several such patterns typically exist, and BEEP simply uses the first one
that the SAT solver returns (although a different BEEP implementation could test
multiple patterns to help identify low-probability errors).
Figure~\ref{fig:beep_steps} \circled{1} illustrates how such a test pattern
appears physically \mpv{within the cells of a codeword}: the target cell is
\texttt{CHARGED}, its neighbors are \texttt{DISCHARGED}, and the SAT solver
freely determines the \mpv{states of the remaining cells} to increase the
likelihood of a miscorrection if the target cell fails. If the SAT solver
\mpv{fails to find such a test pattern, BEEP attempts to craft a pattern using
constraint 2 alone, which, unlike constraint 1, is} essential to observing
miscorrections. Failing that, BEEP simply skips the bit until more error-prone
cells are identified that could facilitate causing miscorrections. We evaluate
\mpi{how successfully BEEP identifies errors} in
Section~\ref{subsubsec:evaluating_beep}, finding that a second pass over the
codeword helps in cases of few or low-probability errors.}

\subsubsection{\jkz{Experimental Testing with Crafted Patterns}} 
\label{subsubsec:running_beep_test_patterns}

\mpiii{BEEP tests a pattern by writing \mpf{it} to the target ECC word, inducing
errors by lengthening the refresh window, and reading out the post-correction
data. Figure~\ref{fig:beep_steps} \circled{2} shows examples of post-correction
error patterns that might be observed during an experiment. Each miscorrection
indicates that an uncorrectable number of pre-correction errors exists, and BEEP
uses the parity-check matrix $\textbf{H}$ to calculate their precise locations.
This is possible because each miscorrection reveals an error syndrome
$\mathbf{s}$ for the (unknown) erroneous pre-correction codeword $\mathbf{c'}$
that caused the miscorrection}. Therefore, we can directly solve for
$\mathbf{c'}$ as shown in Equation~\ref{eqn:system}.
\begin{equation}
\mathbf{s} = \mathbf{H} * \mathbf{c'} = 
c'_0 \cdot \mathbf{H_{\ast,0}} + 
c'_1 \cdot \mathbf{H_{\ast,1}} + 
... + 
c'_n \cdot \mathbf{H_{\ast,n}} \label{eqn:system}
\end{equation}
This is a system of equations with \jkz{one equation for each of $n-k$
unknowns,} \mpi{i.e., one each for the} $n-k$ inaccessible parity bits. There is
guaranteed to be exactly one solution for $\mathbf{c'}$ since the parity-check
matrix always has full rank (i.e., $\mathrm{rank}(\mathbf{H}) = n - k$). Since
we also know the original codeword
($\mathbf{c}=F_{encode}(\mathbf{d})=\mathbf{G}\cdot\mathbf{d}$), we can simply
compare the two (i.e., $\mathbf{c} \oplus \mathbf{c'}$) to determine the
\emph{bit-exact error pattern} that led to the observed miscorrection.
\mpi{Figure~\ref{fig:beep_steps} \circled{3} shows how BEEP updates a list of
learned pre-correction error locations, which the SAT solver then uses to
construct test patterns for subsequent bits. Once all bits are tested, the list
of pre-correction errors yields the number and bit-locations of all identified
error-prone cells.}

\subsubsection{Evaluating BEEP's Success Rate}
\label{subsubsec:evaluating_beep}

To understand how BEEP performs in practice, we evaluate its \emph{success
rate}, i.e., the likelihood that BEEP correctly identifies errors within a
codeword. We use a modified version of EINSim~\cite{eccsimgithub} to perform
Monte-Carlo simulation across 100 codewords per measurement. To keep our
analysis independent of any particular bit-error rate model, we subdivide
experiments by the number of errors ($N$) injected per codeword. In this way, we
can flexibly evaluate the success rate for a specific error distribution using
the law of total probability over the $N$s.

\begin{figure}[h]
    \centering
    \includegraphics[width=\linewidth]{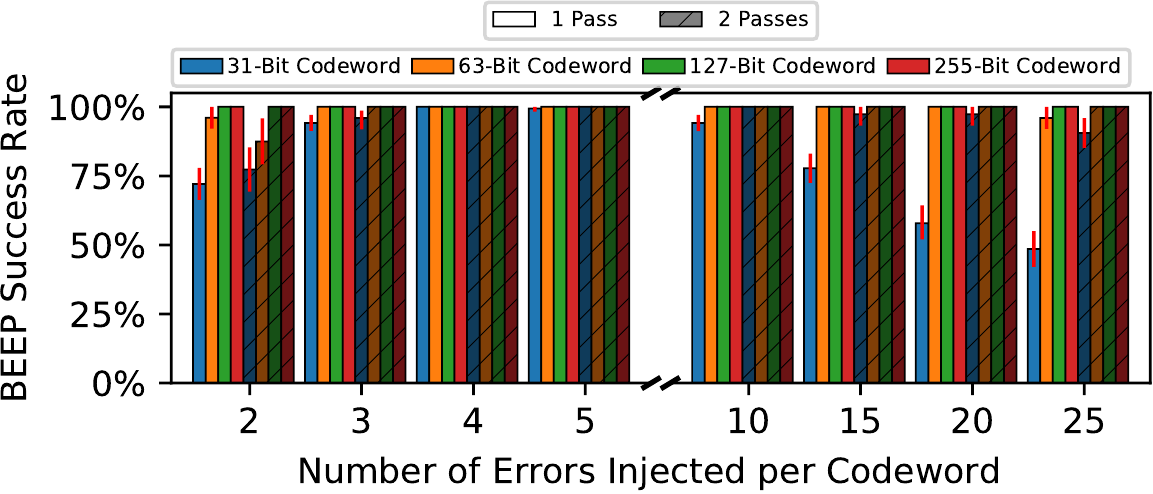}
    \caption{BEEP success rate for 1 vs. 2 passes and
    different codeword lengths and numbers of errors injected.}
    \label{fig:beep_success_rate_npasses}
\end{figure}

\textbf{Number of Passes.} Figure~\ref{fig:beep_success_rate_npasses} shows
BEEP's success rate when using one and two passes over the codeword for
different codeword lengths. \mpii{Each bar shows the median value over the 100
codewords with an error bar showing the 5th and 95th percentiles.} The data
shows that BEEP is highly successful across all tested error counts, especially
for longer 127- and 255-bit codewords that show a 100\% success rate \emph{even
with a single pass}. Longer codewords perform better in part because BEEP uses
one test pattern per bit, which means that longer codes \jkz{lead to} more
patterns. However, longer codewords perform better even with comparable
test-pattern counts (e.g., 2 passes with 31-bit vs 1 pass with 63-bit codewords)
because longer codewords \mpi{simply have more bits (and therefore, error
syndromes) for the SAT solver to consider when crafting a miscorrection-prone
test pattern. On the other hand, miscorrection-prone test patterns are more
difficult to construct for shorter codes that provide fewer bits to work with,
so BEEP fails more often when testing shorter codes.}

\begin{figure}[h]
    \centering
    \includegraphics[width=\linewidth]{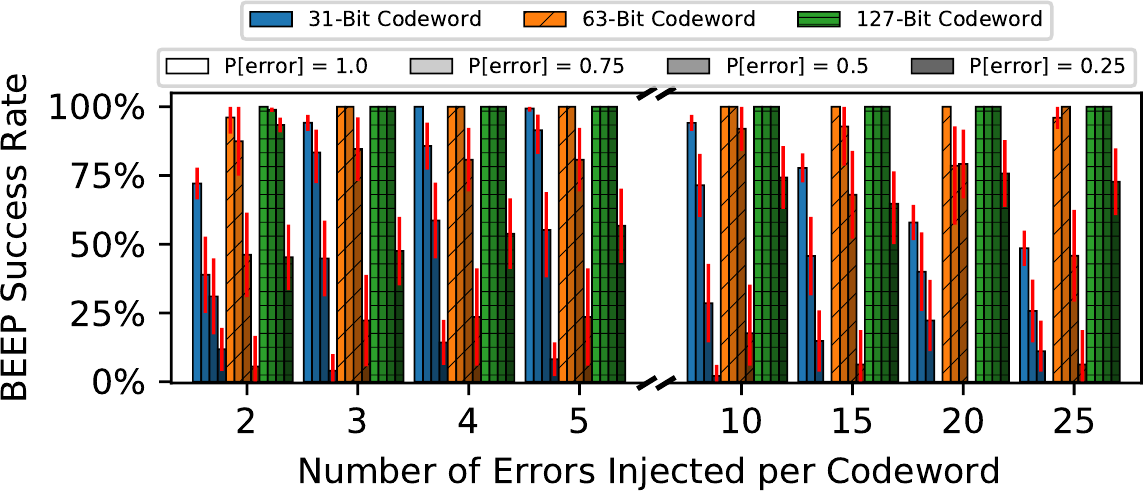}
    \caption{BEEP success rate for different single-bit error
    probabilities using different ECC codeword lengths for different numbers of
    errors injected in the codeword.}
    \label{fig:beep_success_rate_perror}
\end{figure}

\textbf{Per-Bit Error Probabilities.}
Figure~\ref{fig:beep_success_rate_perror} shows how \mpii{BEEP's success rate
changes using a single pass} when the injected errors have different per-bit
probabilities of error (P[error]). This experiment represents a more realistic
scenario where some DRAM cells probabilistically experience data-retention
errors. We see that BEEP remains effective (i.e., \mpii{has a near-100\% success
rate}) for realistic 63- and 127-bit codeword lengths, especially at higher
bit-error probabilities and error counts. \jkz{BEEP generally has a higher
success rate with longer codes compared to shorter ones,} and for shorter
codewords at low error probabilities, the data shows that \jkz{BEEP may require}
more test patterns (e.g., multiple passes) to reliably identify all errors.

It is important to note that, while evaluating low error probabilities is
demonstrative, it \mpiii{represents a pessimistic scenario since a real DRAM
chip exhibits} a mix of low and high per-bit error probabilities.\footnote{Patel
et al.~\cite{patel2017reaper} provide a preliminary exploration of \mpii{how
per-bit error probabilities are distributed throughout a DRAM chip}, but
\mpii{formulating a detailed error model for accurate simulation} is beyond the
scope of \mpiii{our} work.} \mpi{Although \emph{any} error-profiling mechanism
that identifies errors based on when they manifest might} miss low-probability
errors,\footnote{Patel et al.~\cite{patel2017reaper} increase error coverage by
exacerbating the bit-error probability, and their approach \mpii{(REAPER)} can
be used alongside BEEP to help identify low-probability errors.} \jkz{the data
shows that BEEP is resilient} to low error probabilities, especially for longer,
more realistic codewords. Therefore, our evaluations demonstrate that BEEP
effectively enables a new profiling methodology that \mpiii{uses the ECC
function determined by BEER to infer pre-correction errors from observed
post-correction error patterns}.

\subsubsection{\revmp{Other DRAM Error Mechanisms}}
\label{subsubsec:beep_other_error_mechs}

\cro{Although we demonstrate BEEP solely for data-retention errors, BEEP can
potentially be extended to identify errors that occur due to other DRAM error
mechanisms (e.g., stuck-at faults, circuit timing failures). However,
simultaneously diagnosing multiple error models is a very difficult problem
since different types of faults can be nearly indistinguishable (e.g.,
data-retention errors and stuck-at-\texttt{DISCHARGED} errors). Profiling for
arbitrary error types is a separate problem from what we tackle in this work,
and we intend BEEP as a simple, intuitive demonstration of how knowing the ECC
function is practically useful. Therefore, we leave extending BEEP to
alternative DRAM error mechanisms to future work.}

\subsection{Other Use-Cases that Benefit from BEER} 
\label{subsection:use_cases}

\mpii{We identify four additional use cases for which BEER mitigates on-die
ECC's interference with third-party studies by revealing the full ECC function
(i.e., its parity-check matrix).}

\subsubsection{\mpii{Combining Error Mitigation Mechanisms}}

\revmp{If the on-die ECC function is known, a system architect can design a
\mpii{second level of error mitigation} (e.g., rank-level ECC) that better suits
the error characteristics of a DRAM chip with on-die ECC. Figure 1 provides a
simple example of how different ECC functions cause different data bits to be
more error-prone even though the pre-correction errors are uniformly
distributed. This means that on-die ECC \emph{changes} the DRAM chip's
\mpii{software-visible} error characteristics in a way that depends on the
particular ECC function it employs. If the on-die ECC function is known, we can
calculate the expected post-correction error
characteristics\footnote{\revmp{\mpii{By assuming} a given data value
distribution, e.g., \mpii{fixed} values for a \mpii{predictable software}
application, uniform-random data for a general system.}} and build an error
model that accounts for the transformative effects of on-die ECC. Using this
error model, the system architect can make an informed decision when selecting a
secondary mitigation mechanism to complement on-die ECC. For example, architects
could modify a traditional rank-level ECC scheme to asymmetrically protect
certain data bits that are more prone to errors than others as a result of
on-die ECC's behavior~\cite{kraft2018improving, wen2013cd}. In general, BEER
enables system designers to better design secondary error-mitigation mechanisms
to suit the expected DRAM reliability characteristics, thereby improving overall
system reliability.}

\subsubsection{Crafting Targeted Test Patterns}

Several DRAM error mechanisms are highly pattern sensitive, including
RowHammer~\cite{kim2014flipping, mutlu2017rowhammer, mutlu2019rowhammer,
kim2020revisiting}, data-retention~\cite{patel2017reaper, liu2013experimental,
liu2012raidr, khan2014efficacy, khan2016parbor, khan2017detecting,
hamamoto1998retention, kim2009new}, and
reduced-access-latency~\cite{kim2018solar, lee2017design, lee2015adaptive,
chang2016understanding, chang2017understanding}. Different test patterns affect
error rates by orders of magnitude~\hht{\cite{lanteigne2016how,
liu2013experimental, patel2017reaper, kim2020revisiting, khan2017detecting,
khan2016parbor, khan2016case}} because each pattern
exercises different static and dynamic circuit-level effects.  \cro{Therefore,
test patterns are typically designed carefully to induce the worst-case circuit
conditions for the error mechanism under test (e.g., marching
`1's~\cite{adams2002high, liu2013experimental, patel2017reaper,
hassan2017softmc, mrozek2019multi}).} As
Section~\ref{subsubsec:beep_test_patterns} discusses in greater detail, on-die
ECC restricts the possible test patterns to only the ECC function's codewords.
Fortunately, the SAT-solver-based approach that BEEP uses to craft test
patterns generalizes to crafting targeted test patterns for these error
mechanisms also.

\subsubsection{\mpiv{Studying Spatial Error Distributions}}

Numerous prior works~\hht{\cite{shirley2014copula, kim2014flipping,
chang2016understanding, park2016statistical, lee2017design, kim2018solar,
chang2017understanding}}
experimentally study the spatial distributions of errors throughout the DRAM
chip in order to gain insight into how the chip operates and how its
performance, energy, and/or reliability can be improved. These studies rely on
inducing errors at relatively high error rates so that many errors occur that
can leak information about a device's underlying structure. With on-die ECC,
studying spatial error distributions requires identifying
\emph{pre-correction} errors throughout the codeword, including within the
inaccessible parity bits. BEEP demonstrates one possible concrete way by which
BEER \mpf{enables these studies for chips with on-die ECC}.

\subsubsection{\mpiv{Diagnosing Post-Correction Errors}}

A third-party tester may want to determine the physical reason(s) behind an
observed error. For example, a system integrator who is validating a DRAM
chip's worst-case operating conditions may observe unexpected errors due to an
unforeseen defect (e.g., at a precise DQ-pin position). Unfortunately, on-die
ECC obscures both the number and locations of pre-correction errors, so the
observed errors no longer provide insight into the underlying physical error
mechanism responsible. Using BEEP, such errors can be more easily diagnosed
because the revealed pre-correction errors directly result from the error
mechanism.

\subsection{Extensions and Future Work}
\label{subsec:future_work}

\cro{Our work demonstrates that on-die ECC is not an insurmountable problem for
third-party system design and testing. To further explore how tools like BEER
can help clarify a DRAM chip's core reliability characteristics, we identify
several ways in which future studies can build upon our work. We believe these
are promising directions to explore and will further facilitate studying the
reliability characteristics of current and future devices with on-die ECC.}

\noindent
\textbf{Extension to Other Devices.} BEER theoretically applies to any memory
device that uses a linear block code in which we can exploit data-dependent
errors (e.g., \texttt{CHARGED}-to-\texttt{DISCHARGED}) to \mpiv{control which}
miscorrections \mpiv{occur}. A concrete example is DRAM with rank-level ECC,
where BEER can be applied \mpiv{as is}.\footnote{There may be no need to infer
error syndromes from miscorrections if the CPU directly exposes
them~\cite{cojocar19exploiting}.} However, BEER may be extensible to other
memory devices (e.g., Flash memory~\mpiv{\cite{cai2017error, cai2012error,
cai2013error, luo2018heatwatch, luo2018improving, cai2018errors,
meza2015large}}, STT-MRAM~\cite{ishigaki2010multi, zhang2012multi,
kultursay2013evaluating}, PCM~\cite{qureshi2009scalable, lee2009architecting,
wong2010phase, seong2013tri}, Racetrack~\cite{zhang2015hi, parkin2015memory},
RRAM~\mpiv{\cite{wong2012metal, pal2019design, wang2015theory}}) if its core
principles can be adapted for their error models and ECC functions. These
memories all exhibit reliability challenges that BEER can help third-party
scientists and engineers better tackle and overcome.

\noindent
\textbf{Further Constraining the SAT Problem.} 
We believe there are several ways to further constrain the SAT problem,
including (i) prioritizing more likely hardware \mpii{ECC} implementations, (ii)
adding additional SAT constraints for obvious or trivial cases, and (iii)
further constraining the set of test patterns. 

\noindent
\textbf{Improving SAT Solver Efficiency.} Our implementations of BEER and BEEP
express ECC arithmetic (e.g., GF(2) matrix operations, SAT constraints) using
simple Boolean logic equations. An optimized implementation that leverages
native GF(2) BLAS libraries (e.g., LELA~\cite{hovinen2011lela}) and advanced SAT
solver theories (e.g., SMT bitvectors~\cite{brummayer2009boolector}) could
drastically improve BEER's performance, enabling BEER for a wider variety of ECC
functions. Taking this a step further, future work could reformulate BEER's SAT
problem mathematically in order to directly solve for the parity-check matrix
that can produce a given miscorrection profile. Such \mpii{an approach could
identify the solution significantly faster than using a SAT solver to  
perform a brute-force exploration of the entire solution space.}

%% file: 8_related_work.tex
\section{Related Work}
\label{sec:related}

\cro{To our knowledge, this is the first work to (i) determine the full on-die
ECC function and (ii) recover the number and bit-exact error locations of
pre-correction errors in DRAM chips with on-die ECC without any insight into the
ECC mechanism or any hardware modification. We distinguish BEER from related
works that study on-die ECC, techniques for reverse-engineering DRAM ECC
functions, and DRAM error profiling.}

\noindent
\textbf{On-Die ECC.} Several works study on-die ECC~\cite{nair2016xed,
patel2019understanding, cha2017defect, gong2018duo}, but only Patel et
al.~\cite{patel2019understanding} attempt to identify pre-correction error
characteristics without bypassing or modifying the on-die ECC mechanism.
Although \jkz{Patel et al.~\cite{patel2019understanding}} statistically infer
high-level characteristics about the ECC mechanism and pre-correction errors,
their approach has several \emph{key limitations} (discussed in
Section~\ref{sec:intro}). BEER overcomes these limitations and identifies (1)
the full ECC function and (2) the bit-exact locations of pre-correction errors
without requiring any prerequisite knowledge about the errors being studied.

\noindent
\textbf{Determining ECC Functions.} Prior works reverse-engineer ECC
characteristics in Flash memories~\cite{van2015mathematical, van2017bit,
wise2018reverse}, DRAM with rank-level ECC~\cite{cojocar19exploiting}, and
on-die ECC~\cite{patel2019understanding}. However, none of these works can
identify the full ECC function \mpi{by studying data only at the external DRAM
chip interface because they either require (1) examining the} encoded
data~\cite{van2015mathematical, van2017bit, wise2018reverse}, \mpi{ (2)}
injecting errors directly into the codeword~\cite{cojocar19exploiting}, or
\jkz{(3) knowing when} an ECC correction is performed and \mpii{obtaining} the
resulting error syndrome~\cite{cojocar19exploiting}.  On-die ECC provides
\emph{no insight} into the error-correction process and does not report if
\mpi{or} when a correction is performed.

\noindent
\textbf{DRAM Error Profiling.} Prior work proposes many DRAM error profiling
\jkx{methodologies~\cite{hamamoto1995well, hamamoto1998retention,
shirley2014copula, jung2015omitting, weis2015thermal, weis2015retention,
jung2014optimized, khan2016parbor, khan2016case, khan2014efficacy,
hassan2017softmc, lee2017design, kim2018solar, kim2019d, gao2019computedram,
chang2017understanding, venkatesan2006retention, liu2012raidr,
liu2013experimental, qureshi2015avatar, patel2017reaper, kim2018dram,
lee2015adaptive, kim2014flipping, chang2016understanding, kim2020revisiting,
cojocar19exploiting, kraft2018improving, patel2019understanding}}.
Unfortunately, none of these approaches are capable of identifying
pre-correction error locations throughout the entire codeword (i.e., including
within parity-check bits).

%% file: 9_conclusion.tex
\section{Conclusion}

We introduce Bit-Exact Error Recovery (BEER), a new methodology for determining
the full \mpii{DRAM} on-die ECC function (i.e., its parity-check matrix) without
requiring hardware support, \mpi{prerequisite knowledge about the DRAM chip or
on-die ECC mechanism,} or access to \mpf{ECC metadata (e.g., parity-check bits,
error syndromes)}. We use BEER to determine the on-die ECC functions of 80 real
LPDDR4 DRAM chips and show that \mpi{BEER} is both effective and practical using
rigorous simulations. We discuss five concrete use-cases for BEER, including
BEEP, a new DRAM error profiling methodology capable of inferring \mpi{exact}
pre-correction error counts and locations. We believe that BEER takes an
important step towards enabling effective third-party design and testing around
DRAM chips with on-die ECC and are hopeful that BEER will enable many new
studies \mpiii{going forward}.